\documentclass{ieeeaccess}
\usepackage{cite}
\usepackage{amsmath,amssymb,amsfonts}
\usepackage{graphicx}
\usepackage{textcomp}
\usepackage{color, colortbl}
\usepackage{multirow}% http://ctan.org/pkg/multirow
\usepackage{hhline}% http://ctan.org/pkg/hhline
\usepackage{booktabs}
\usepackage{tabularx,ragged2e}
\usepackage{supportcaption}
\usepackage{subcaption}
\usepackage{enumerate}
\usepackage{placeins}
\usepackage{footnote}
\usepackage[english]{babel}
\usepackage[bottom]{footmisc}
\usepackage{url}

\usepackage{color, colortbl}
\definecolor{Gray}{gray}{0.9}

\def\BibTeX{{\rm B\kern-.05em{\sc i\kern-.025em b}\kern-.08em
    T\kern-.1667em\lower.7ex\hbox{E}\kern-.125emX}}

\begin{document}
\history{Date of publication xxxx 00, 0000, date of current version xxxx 00, 0000.}
\doi{10.1109/ACCESS.2020.DOI}

\title{Distributed Vehicular Computing at the Dawn of 5G: A Survey}

\author{\uppercase{Ahmad ALHILAL}\authorrefmark{1},
\IEEEmembership{Member, IEEE},
\uppercase{Benjamin~FINLEY}\authorrefmark{2},
\uppercase{Tristan~BRAUD}\authorrefmark{3}, \IEEEmembership{Member, IEEE},
\uppercase{Dongzhe~Su}\authorrefmark{4},
and \uppercase{Pan~HUI}\authorrefmark{2,5,6},
\IEEEmembership{Fellow, IEEE}.}
\address[1]{Department
of Computer Science Engineering, The Hong Kong University of Science and Technology, Hong Kong}
\address[2]{Department of Computer Science, University of Helsinki, Helsinki, Finland}
\address[3]{Division of Integrative Systems and Design, The Hong Kong University of Science and Technology, Hong Kong}
\address[4]{The Hong Kong Applied Science and Technology Research Institute (ASTRI), Hong Kong}
\address[5]{Computational Media and Arts Thrust Area, The Hong Kong University of Science and Technology (Guangzhou), China}
\address[6]{Division of Emerging Interdisciplinary Area, The Hong Kong University of Science and Technology, Hong Kong}

\markboth
{Distributed Vehicular Computing at the Dawn of 5G}
{A. Alhilal \headeretal: Distributed Vehicular Computing}
\corresp{Corresponding author: Benjamin Finley (e-mail: benjamin.finley@helsinki.fi).}

\begin{abstract}
Recent advances in information technology have revolutionized the automotive industry, paving the way for next-generation smart vehicular mobility. Vehicles, roadside units, and other road users can collaborate to deliver novel services and applications. These services and applications require 1) massive volumes of heterogeneous and continuous data to perceive the environment, 2) reliable and low-latency communication networks, 3) real-time data processing that provides decision support under application-specific constraints. Addressing such constraints introduces significant challenges for current communication and computing technologies. Relatedly, the fifth generation of cellular networks (5G) was developed to respond to communication challenges by providing for low-latency, high-reliability, and high bandwidth communications. As a major part of 5G, edge computing allows data offloading and computation at the edge of the network, ensuring low-latency and context-awareness, and 5G efficiency. In this work, we aim at providing a comprehensive overview of the state of research on vehicular computing in the emerging age of 5G and big data.
\end{abstract}


\titlepgskip=-15pt

\maketitle

\section{Introduction}
\label{sec:intro}

The automotive industry is on the verge of one of the most dramatic paradigm shifts in its history. 
An increasing number of vehicles embed sensing, computation, and wireless communication capabilities. Such vehicles feature onboard units (OBU), global positioning system (GPS) units, onboard radio modules, such as IEEE 802.11p, long-term evolution (LTE), or 5G modules, and other onboard units. These units perceive the surrounding environment and perform computation and communication.
Similar to vehicles, the road infrastructure itself is also embedding more intelligence. Induction loop detectors can detect vehicles passing or arriving at a certain location such as approaching a traffic light or in motorway traffic. Specifically, the pavement is equipped with an insulated, electrically conducting loop which detects the presence of vehicles and can be connected to roadside units (RSUs).
Roadside units (RSUs) are transceivers mounted along a road or pedestrian passageway to interact with vehicles and perform computation, communication, and storage tasks. These capabilities enable the vehicles and the infrastructure to form a vehicular ad-hoc network (VANET) spontaneously and without any additional infrastructure~\cite{hartenstein2009vanet}. However, due to the mobility of the vehicles and dynamic nature of traffic, links change frequently leading to ever-changing topology and network partitioning. Resultingly, sparse and heavy traffic frequently alternate leading to intermittent connectivity and network congestion. These dramatic changes introduce high latency variability which impacts the quality of service. These conditions complicate the deployment of vehicular applications that require real-time interactions and prevent the deployment of time-critical safety applications.

\begin{figure}[!t]
    \centering
    \includegraphics[width=0.9\linewidth]{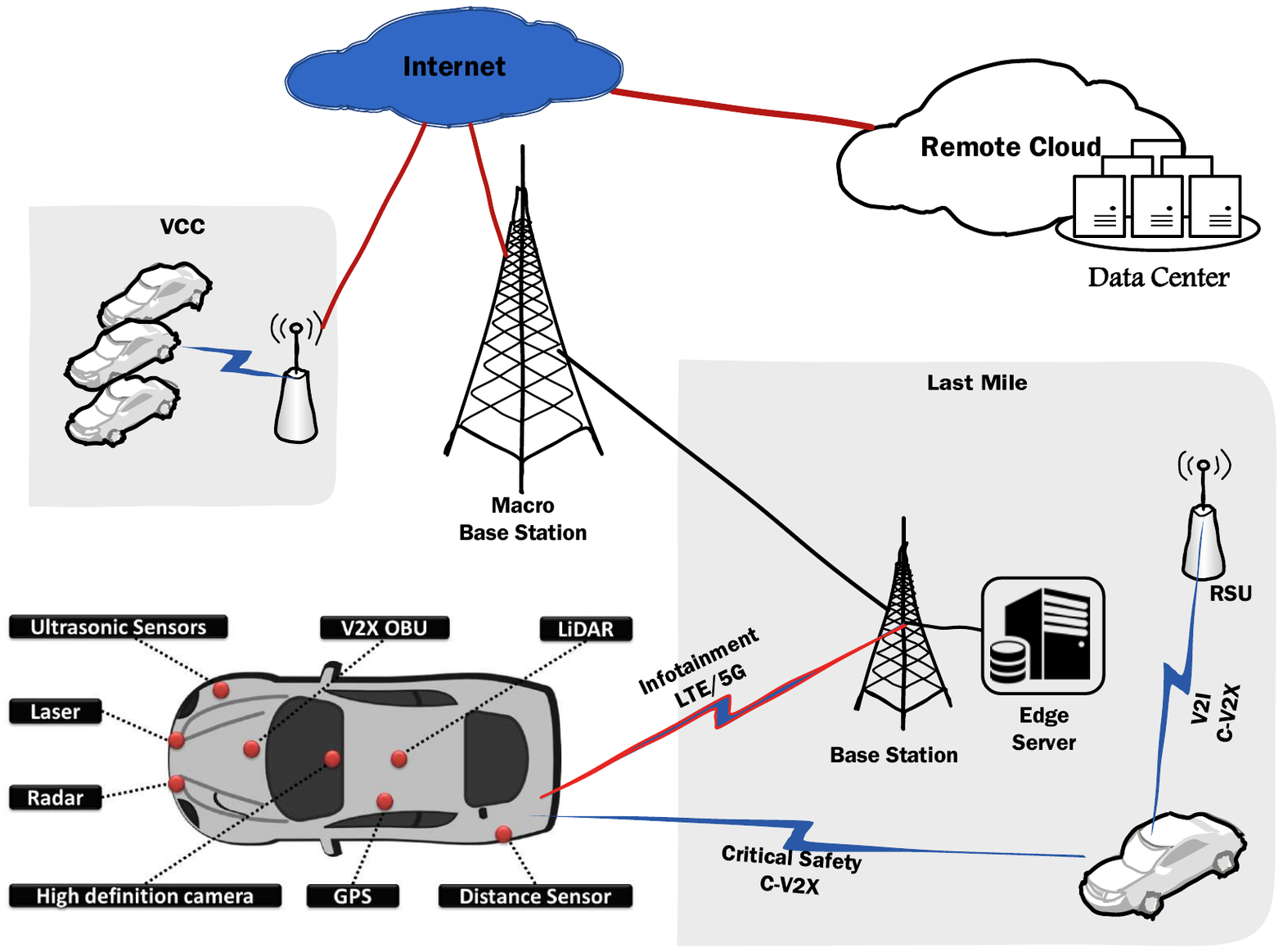}
    \caption{Overview of on-board units~\cite{hamida2015security} and computation and communication resources for two vehicular applications. The applications leverage vehicular edge computing (VEC) and vehicular cloud computing (VCC) for computation resources, mobile network for internet access, and C-V2X direct communication for critical safety communication.}
    \label{fig:VANEToverview}
\end{figure}

The development of connected vehicular systems paves the way for new services and business opportunities. The deployment of the technologies, infrastructure, and services relies on an interdisciplinary effort involving not only manufacturers, but also network operators, service providers, and governmental authorities. Network operators provide network access, whereas service providers provide access to specific services and bill subscribed users~\cite{moustafa2009vehicular}. For example, service providers can collect real-time traffic data, detect traffic congestion, and disseminate such information to vehicles, either through RSUs or cellular communication~\cite{jayapal2016road}. Finally, government authorities play a critical role in aligning all actors towards providing safe, reliable, and interoperable road services. 
Such a collaboration opens up numerous possibilities for potential life-changing applications in the area of Intelligent Transportation Systems (ITS). These applications range from critical safety-related applications and driver assistance systems to location-based and traffic management services. The computation and communication resources involved in vehicular applications vary based on the application requirements, as shown in Figure~\ref{fig:VANEToverview}. At one end of the spectrum, infotainment applications access the internet through mobile networks (LTE and 5G) thus allowing multimedia streaming or web browsing. On the other end, safety-critical applications rely primarily on vehicles, clouds, and edge servers for decision-making. These applications also tend to exploit reliable and lower-latency communication solutions, for example, cellular vehicle-to-everything (C-V2X). C-V2X enables vehicle to vehicle (V2V), vehicle to infrastructure (V2I), and other communication variants to address the real-time constraints.

Furthermore, road elements (traffic lights, lampposts, induction loop detectors, RSUs) and road users (vehicles, pedestrians' smartphones) produce traffic data resulting in significant data heterogeneity and volume due to sensor ubiquity and diversity. Thus processing heterogeneous and big data streams becomes another challenge. Additionally, as previously stated, the connection may be intermittent, resulting in massive data bursts that need to be processed in real-time. Analyzing this data and promptly extracting meaningful and useful information requires the consideration of specific big data architectures in the deployment of connected vehicular systems~\cite{alhilalcad3,amini2017big}.

Cloud computing (CC) enables ubiquitous, convenient, and on-demand access to a shared pool of configurable computing resources (e.g., networks, servers, storage, applications, and services)~\cite{mell2011nist}. However, many applications have strict latency requirements, which makes edge computing (EC) a better candidate compared to remote centralized clouds. EC promises to deliver scalable, highly responsive cloud services for mobile computing and masks transient cloud outages. In contrast to CC, EC features proximity to the subscribing vehicles, context-awareness, dense geographical distribution, and support for mobility~\cite{zhang2017mobile,satyanarayanan2017emergence}. For instance, EC-assisted traffic management enables monitoring the lane occupancy using traffic data and changing traffic signal phases accordingly.

There is a growing volume of research related to vehicular communication and computing. However, as far as we are aware, the domain lacks a systematic survey that studies the existing literature on distributed vehicular computing, pinpoints the applications and requirements, highlights the methodologies, and determines the enabling technologies. 
Therefore in this work, we provide a survey that fulfills these comprehensive goals. We first focus on the communication methods and evaluate how current and future technologies can assist vehicular networking in handling significant network load while keeping network latency at a minimum. We then move on to potential data offloading and computing architectures, ranging from road to city scale. As a promising paradigm, we discuss edge computing methods including fog computing (using RSUs) and multiaccess edge computing (MEC), a standardized ETSI architecture. Afterwards, we review different data analytics algorithms to support decision-making based on large volumes of heterogeneous data. Finally, we discuss insights and open issues, among them future communication, background services, and safety applications for further research and development of novel applications.

\begin{table}[!t]
  \begin{center}
    \caption{List of acronyms}
    \label{tab:acrynoms}
    \footnotesize 
    \begin{tabular}{p{1.5cm}p{6cm}}
    
    \toprule % <-- Toprule here
    \textbf{Acronym} & \textbf{Definition}   \\ %& \textbf{Description} 
    
    \toprule % <-- Toprule here
    \rowcolor{Gray}
    V2X  & Vehicle-to-Everything  \\
    V2I  &   Vehicle-to-Infrastructure \\ 
    V2V  &   Vehicle-to-Vehicle \\ 
    C-V2X  & Cellular (LTE)-based Vehicle-to-Everything  \\
    NR C-V2X  & Cellular (5G)-based Vehicle-to-Everything  \\
    \toprule % <-- Toprule here
    \rowcolor{Gray}
    ITS  &   Intelligent Transportation System \\
    
    OBU  & On-Board Unit  \\ 
 
    RSU  & Road Side Unit  \\ 

    CVs & Connected Vehicles    \\ 

    AV/D & Autonomous Vehicle/Driving     \\ 

    ADAS  & Advanced Driver Assistance Systems   \\ 
   
    \toprule % <-- Toprule here
    \rowcolor{Gray}
    DSRC  & Dedicated Short Range Communication  \\ 
    WAVE  &   Wireless Access in Vehicular Environment \\ 
    CAM  &   Cooperative  Awareness  Message \\
    DENM  &   Decentralized Environmental Notification Message \\ 
    WSMP  &   WAVE Short Message Protocol \\ 
          
    \toprule % <-- Toprule here
    \rowcolor{Gray}
    3GPP & The 3rd Generation Partnership Project\\
    LTE  & Long-Term Evolution  \\ 
    BM-SC  &    Broadcast  Multicast  Service  Center  \\ 
    eMBMS  &    evolved  Multimedia  Broadcast  Multicast Service \\ 
    EPC &   Evolved Packet Core  \\ 
    RAN & Radio Access  Network\\
    RAT & Radio Access Technology\\
    
    \toprule % <-- Toprule here
    UE &   User Equipment  \\ 
    PDN  &  Packet Data Network   \\ 
    P-GW  &   PDN Gateway   \\ 
    S-GW  &    Serving  Gateway  \\ 
    MME  &  Mobility  Management  Entity   \\ 
    MBSFN  &  Multimedia Broadcast Multicast Service Single Frequency Network   \\
    
    \toprule % <-- Toprule here
    \rowcolor{Gray}
    5G  & Fifth-Generation Mobile Network    \\
    MIMO  &  Multiple Input Multiple  Output    \\ 
    SDN  & Software Defined Networking    \\ 
    NFV  & Network Function Virtualization    \\ 
    uRLLC & ultra Reliable Low Latency Communication   \\ 
    eMBB  & evolved Mobile Broad Band    \\ 
    mMTC  & massive Machine Type Communication   \\ 
    AMF  & Access and Mobility Management Function   \\ 
    SMF  &  Session  Management  Function \\ 
    UPF  &  User Plan Function \\ 
    KPIs  &  5G Key Performance Indicators \\ 
     
    \toprule % <-- Toprule here
    \rowcolor{Gray}
    CC  & Cloud Computing  \\ 
    EC  & Edge Computing   \\
    MEC  & Multiaccess Edge Computing   \\
    FC  & Fog Computing   \\
   
    \end{tabular}
  \end{center}
\end{table}

%to edit and revise
\noindent The contributions of this survey are threefold:
\begin{enumerate}
    \item \textbf{Challenges and Requirements.} We characterize the vehicular environment including the challenges and requirements of real-world vehicular applications.
    \item \textbf{Bottom-up Overview.} We provide a comprehensive study of existing communication technologies and computation paradigms. We also examine promising communication and computing technologies before further investigating big data analytics frameworks and ML-empowered vehicular applications.
    \item \textbf{Integrated Architecture and Case Studies.} We study the requirements and potential architectures of ITS systems at local, neighborhood, and city scales, and showcase real-world scenarios.  
    \item \textbf{Future direction.} We discuss insights and open issues, which shed light on the development of future novel vehicular applications and services.
\end{enumerate}

\section{Challenges of vehicular Applications}
\label{sec:challenges}
The deployment of vehicular applications will face multiple challenges relating to storage, computing capabilities, network, privacy, and security.

\textbf{Data Volume, Variety, and Velocity (3Vs).} 
Connected vehicles contain a wide variety of sensors that continuously produce large amounts of data. RGB cameras alone generate 20 to 40 Mbps and radar sensors produce between 10 and 100 Kbps. The move from simply connected vehicles to autonomous vehicles will likely further increase sensor data volume (as autonomous vehicles will have more sensors). An autonomous vehicle may include 4-6 radar sensors generating 0.1 - 15 Mbps per sensor, 1-5 lidar sensors generating 20-100 Mbps per sensor, 6-12 RGB camera sensors generating 500 - 3500 Mbps per sensor,  8-16 ultrasonic sensors generating less than 0.01 Mbps per sensor, and vehicle motion, global navigation satellite system (GNSS) and inertial measurement unit (IMU) with <0.1 Mbps per sensor. The total sensor bandwidth is thus between 3 Gbps (10.8 Tb/h) and 40 Gbps (144 Tb/h)~\cite{zhang2019mobile}. Therefore, disseminating such massive data to remote servers or processing with near-real-time constraints are non-trivial issues.

\textbf{Limited Computing Resources:} 
The addition of thousands of new connected devices stresses not only the networks but also the computational power (for example at RSU) ~\cite{lamb2019analysis}. Advanced driver-assistance systems (ADASs) and autonomous vehicles (AVs) with numerous onboard sensors will generate large amounts of data to be processed. Additionally, ensuring a holistic view of the ambient environment is beyond the capacity of a single vehicle, as it requires aggregating the point of view of multiple vehicles to recreate the scene~\cite{golestan2016situation,nwiabu2011situation}. Therefore, off-board computation is crucial to cope with such situations ~\cite{erso2016adas,kukkala2018advanced}. 

\textbf{Rapid Topology Change and High mobility:} 
The relative speed between vehicles ranges from tens of Km/h (vehicles travelling in the same direction on an urban street) to over 280 Km/h (vehicles travelling in opposite directions on a highway). Thus, vehicles may be members of a given VANET for only a very short time leading to rapid and frequent network topology changes~\cite{da2014data,lu2014connected}. Additionally, traffic congestion can be predictable (e.g., due to rush hours) and unpredictable (e.g., due to traffic accidents)~\cite{zhou2019erl}. These phenomena lead to large volumes of data (due to frequent updates) in some urban areas and may run into limitations in VANET-based application scalability~\cite{araniti2013lte}.

\textbf{Detrimental Delay:} 
Propagation and queuing delay are major sources of delay in ITS. Propagation delay depends on the communication medium and the physical distance between the communication source and the destination. For example, if the source and destination are in the same neighbourhood at a distance of 1 km, the propagation delay will be $\approx 5 \mu sec$. However, if they are located in different countries at a distance of 12,000 km the delay can reach $\approx$ 58.2 ms. 

Queuing delay denotes the network delay of data while waiting in network queues to be initially sent by the sender, forwarded by intermediate nodes, or processed by the destination. Queuing delay is related to the number of transmitting vehicles and the volume of data sent by each vehicle (i.e., the network traffic), the available number of links between the source and destination, max queue lengths on the nodes, and the service policy (e.g., critical safety or non-safety) which determines the queuing prioritization~\cite{PEREZNEIRA2009125}. Overall, the networking overhead and latency associated with remote cloud resources could degrade the overall performance and prove detrimental to road safety~\cite{lamb2019analysis,popescu2017characterizing,KHANVILKAR2005401}. Moreover, traffic volume is increasing with user demand and will further heavily burden backhaul links and lead to even longer latencies~\cite{wang2017survey}. 

\textbf{Security and Privacy Concern:}
Some ITS applications and services require vehicles and RSUs to exchange messages containing potentially sensitive information such as real-time locations. This communication takes place over networks that are by design somewhat easily accessible \footnote{Easy accessibility is important because many ITS applications, such as cooperative perception, demonstrate local network effects (wherein the benefit for each user scales with total number of local users).} thus prompting security and privacy challenges~\cite{qu2015security}. Furthermore, the richer the data sharing the more potential exists for tactics like predatory marketing and user tracking~\cite{acohido2019privacy}. As the number of services grows passengers may also use services over multiple networks, therefore. magnifying the potential for cyber-security problems or privacy violations. 

\section{V2X Communication Overview}
\label{sec:review}
Vehicles and roadside infrastructure use multiple wireless technologies to communicate. The most promising wireless communication technologies can be classified into short-range communications such as dedicated short-range communication (DSRC)~\cite{kenney2011dedicated} and ITS-G5~\cite{festag2014cooperative, ETSI.EN.302.663.V1.2.0}, and long-range communications including long term evolution (LTE) and 5G. These technologies vary according to their range, capacity, and communication latency. Each technology is thus suitable for a specific class of applications. 

\subsection{Short-Range Communication (DSRC and ITS-G5)}
Dedicated Short Range Communication (DSRC) is a vehicular communication technology that operates in 75 MHz of licensed spectrum in the 5.9 GHz band in the United States and supports intelligent transportation systems. \textbf{DSRC} allows vehicles and RSUs to form vehicular ad-hoc networks through vehicle-to-vehicle (V2V) and infrastructure-to-vehicle (V2I) communications. DSRC operates based on the interoperability between standards that form its protocol stack. IEEE 802.11p~\cite{ieee2010.802.11p} is the native technology, a derivative of the IEEE 802.11 (WiFi) standard, at the physical and medium access control layers. In data link, network, and transport layers, DSRC employs a family of IEEE 1609 standards: the IEEE 1609.2~\cite{ieee2016.1609.2}, 1609.3~\cite{ieee2016.1609.3}, and 1609.4~\cite{ieee2016.1609.4} standards for security, network services (including the WAVE short message protocol - WSMP) and multi-channel operation. WSMP is a bandwidth-efficient protocol used for exchanging single-hop messages and non-routed data. WSMP sends packets referred to as WAVE short messages (WSMs). IEEE 802.11p and IEEE 1609 standards allow vehicles to operate in a rapidly varying environment and exchange messages either without having to join a basic service set (BSS) or within a WAVE BSS~\cite{kenney2011dedicated,jiang2008ieee}. DSRC seeks to enable vehicular collision prevention applications that depend on periodic data exchanges among vehicles and between vehicles and roadside infrastructure with strict round trip latency, broadcast frequency, and packet error rate requirements~\cite{machardy2018v2x}. According to ETSI, cooperative collision avoidance requires a guaranteed maximum latency time of 50 ms and a minimum frequency of 10 hz to broadcast pre-crash state in cooperative awareness messages that are associated with direct V2V communication~\cite{etsi2009intelligent}. DSRC fulfills the requirements of such applications while also providing high security, low latency, and high-speed direct communication between entities, without involving a centralized network infrastructure~\cite{kenney2011dedicated}.

\textbf{ITS-G5} is an analogous European technology for vehicular communication that also uses the 5.9 GHz frequency band but is adapted to European requirements. This standard is developed by the European Telecommunications Standards Institute (ETSI) to guarantee interoperability among communication devices from different manufacturers. Similar to DSRC, it carries V2V and V2I in an ad-hoc fashion. They are, however, different only in the way they access the channel~\cite{festag2014cooperative, ETSI.EN.302.663.V1.2.0}. The ITS G5 approach includes a model consisting of state machines and different tunable parameters to control the medium access of all nodes. ITS-G5 standard adds features for decentralized congestion control methods to control the network load~\cite{eckhoff2013performance, ETSI.EN.302.663.V1.2.0}.
 
\subsection{Long-Range Communication (LTE and 5G)}
The \textbf{Long term evolution (LTE)} standard is mobile communications standard developed by the 3rd Generation Partnership Project (3GPP~\footnote{\url{https://www.3gpp.org}}). The LTE system infrastructure comprises a core network, also known as an evolved packet core (EPC), and an access network, referred to as an evolved universal terrestrial radio access network (E-UTRAN). Further details on the basics of the LTE architecture can be found from 3GPP~\cite{lte3gpp}.

In the vehicular context, 3GPP has developed Cellular-V2X (C-V2X) to operate in 5.9 GHz band (similar to DSRC) in addition to the licensed carriers via network infrastructure. This enables direct communications in the absence of cellular infrastructure in a distributed manner~\cite{hakeem20205g}. C-V2X works in two transmission modes to support ITS services: 1) C-V2X/PC5 which supports V2X direct sidelink communications, allowing vehicles and RSUs to inter-communicate directly without the need for infrastructure, and thus providing lower delay, higher throughput, lower energy consumption, and better spectrum utilization~\cite{chen2017vehicle}, 2) C-V2X/Uu communications to connect road users (e.g., vehicles and RSUs) indirectly through LTE infrastructure. In this mode, since the V2X transmissions are scheduled, interference and collisions are lessened~\cite{hakeem20205g}.

\begin{figure*}[!t]
    \centering
    \includegraphics[width=0.9\linewidth]{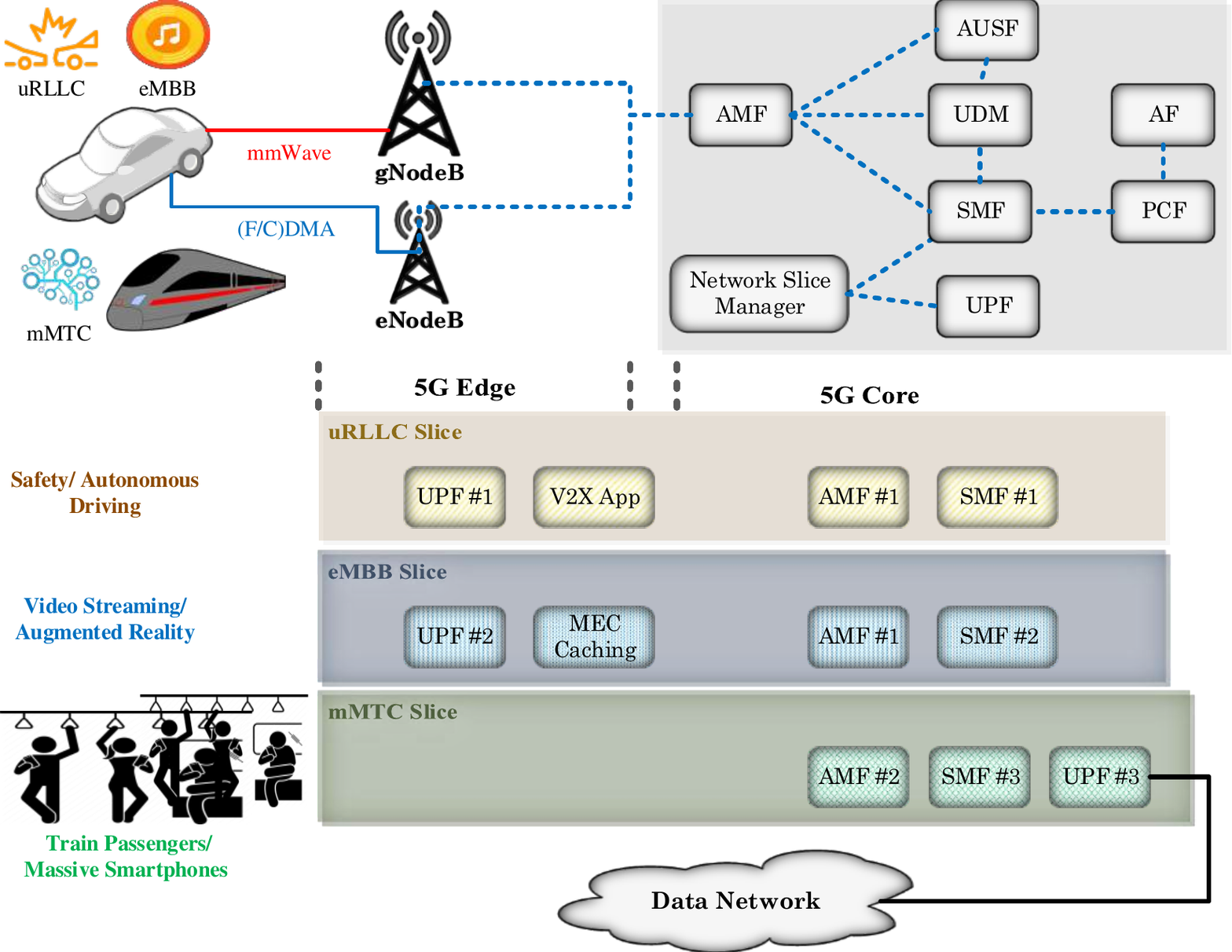}
    \caption{5G network slicing using Network Functions (NFs), Access and Mobility Management (AMF), Authentication Server (AUSF), Unified Data Management (UDM), Session Management (SMF), Application (AF), and Policy Control Functions (PCF)~\cite{storck20195g}. The car leverages two slices, a uRLLC slice to support road safety or autonomous driving, and an eMBB slice for video streaming or augmented reality. While a train leverages an mMTC slice to handle the massive smartphone traffic of the passengers~\cite{5gamericas2018v2x}.}
    \label{fig:5gcran}
\end{figure*}

3GPP has been developing C-V2X to base on the fifth-generation 5G mobile communications standard, leading to New Radio (NR) C-V2X which is going to be compatible with the evolution path of 5G. NR C-V2X is gaining momentum with deployments in many countries. In general, 5G aims for ultra-reliable low-latency communication. Besides, 5G supports ultra-high throughput which is 10-100$\times$ higher compared to LTE. Similar to LTE C-V2X, \textbf{sidelink} mode allows direct communication between vehicles, while \textbf{indirect (via infrastructure)} mode works inside the coverage range of a gNodeB. However, NR C-V2X supports unicast, group cast, or broadcast transmission modes while LTE C-V2X only supports broadcast transmission mode. The 5G basis ensures interoperability with earlier communications systems such as LTE (i.e., Non-standalone) and provides good communication performance for vehicular applications. Additionally, 5G supports integration (beyond simple IP-based connectivity) with many existing communications and telecommunication systems including 3G, 4G, WiFi variants, ZigBee, and Bluetooth. This integration provides vehicular networks with more flexibility, allowing vehicles, drivers, passengers, and pedestrians to leverage the most suitable network for their selected application~\cite{shah20185g}.

In addition, a variety of 5G and 5G-adjacent technologies including software-defined networks (SDN), network function virtualization (NFV), and multiaccess edge computing (MEC) are accelerating the ongoing migration of intelligence closer to the users. These paradigms are the building blocks of the network softwarization happening in mobile networks~\cite{carugi2018key}. \textbf{5G ecosystem} enables vehicle manufacturers, solution integrators, network and service providers, and small and medium-sized enterprises (SMEs) to efficiently compete and cooperate. Within the 5G System, end-to-end network slicing, service-based architecture, software-defined networking (SDN), and network functions virtualisation (NFV) are the fundamental pillars to support the heterogeneous key performance indicators (KPIs) of the new use cases in a cost-efficient way. SMEs will be able to provide technological solutions which comply with the overall system design. Manufacturers and solution integrators can offer rapid deployment enabled by virtualisation and standardised interfaces to assimilate the advent level of innovation. Mobile network operators (MNOs) and infrastructure providers will create tailored slices with specific functionalities and services to address the requirements of vertical industries~\cite{redana20195g}.

\begin{figure*}[!t]
    \centering
    \begin{subfigure}[b]{0.46\textwidth}
        \centering
        \includegraphics[width=\textwidth]{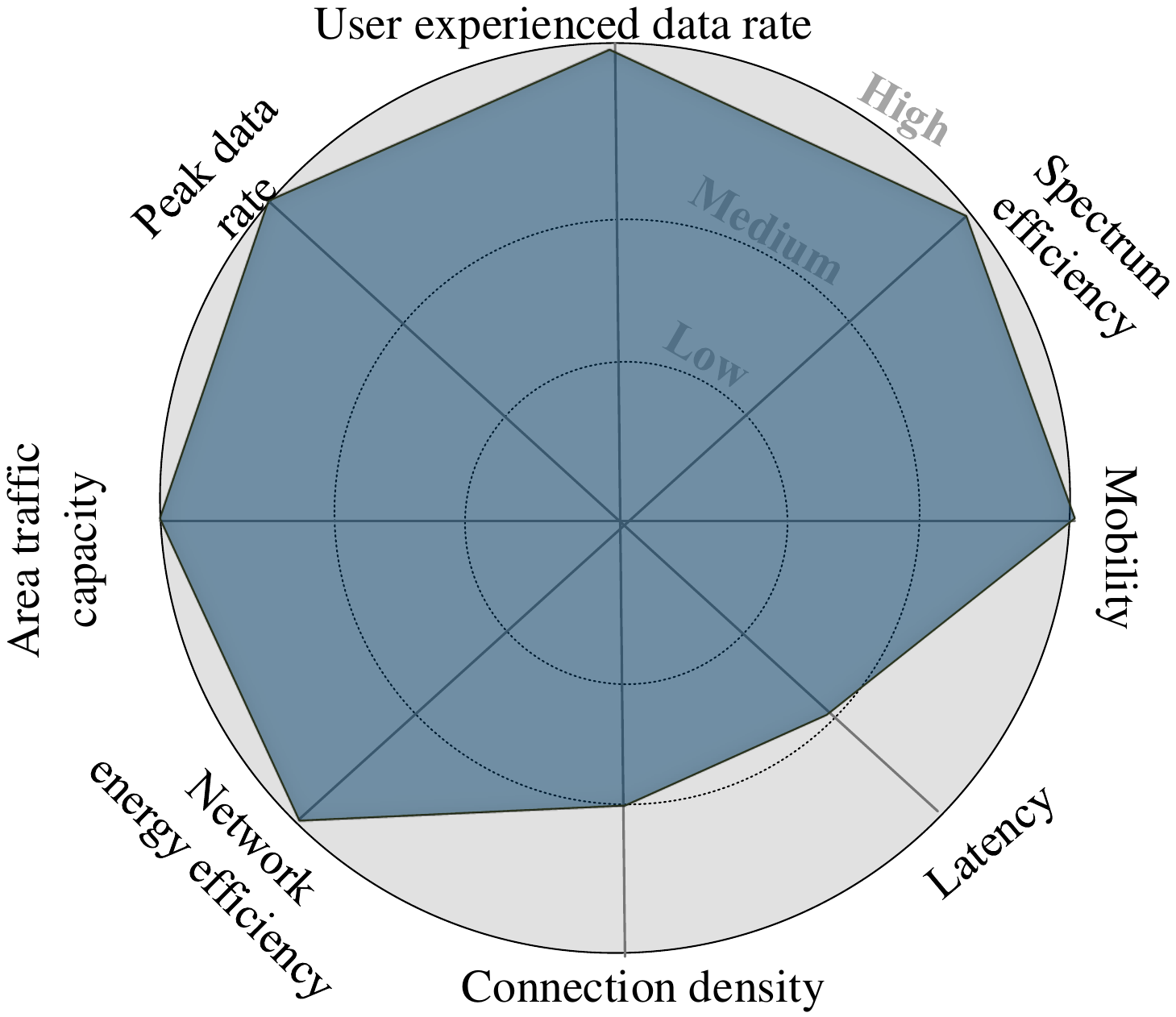}
        \caption{Infotainment applications (e.g. Video streaming, AR)}
        \label{fig:info5G}
    \end{subfigure}
    \hfill
    \begin{subfigure}[b]{0.46\textwidth}
        \centering
        \includegraphics[width=\textwidth]{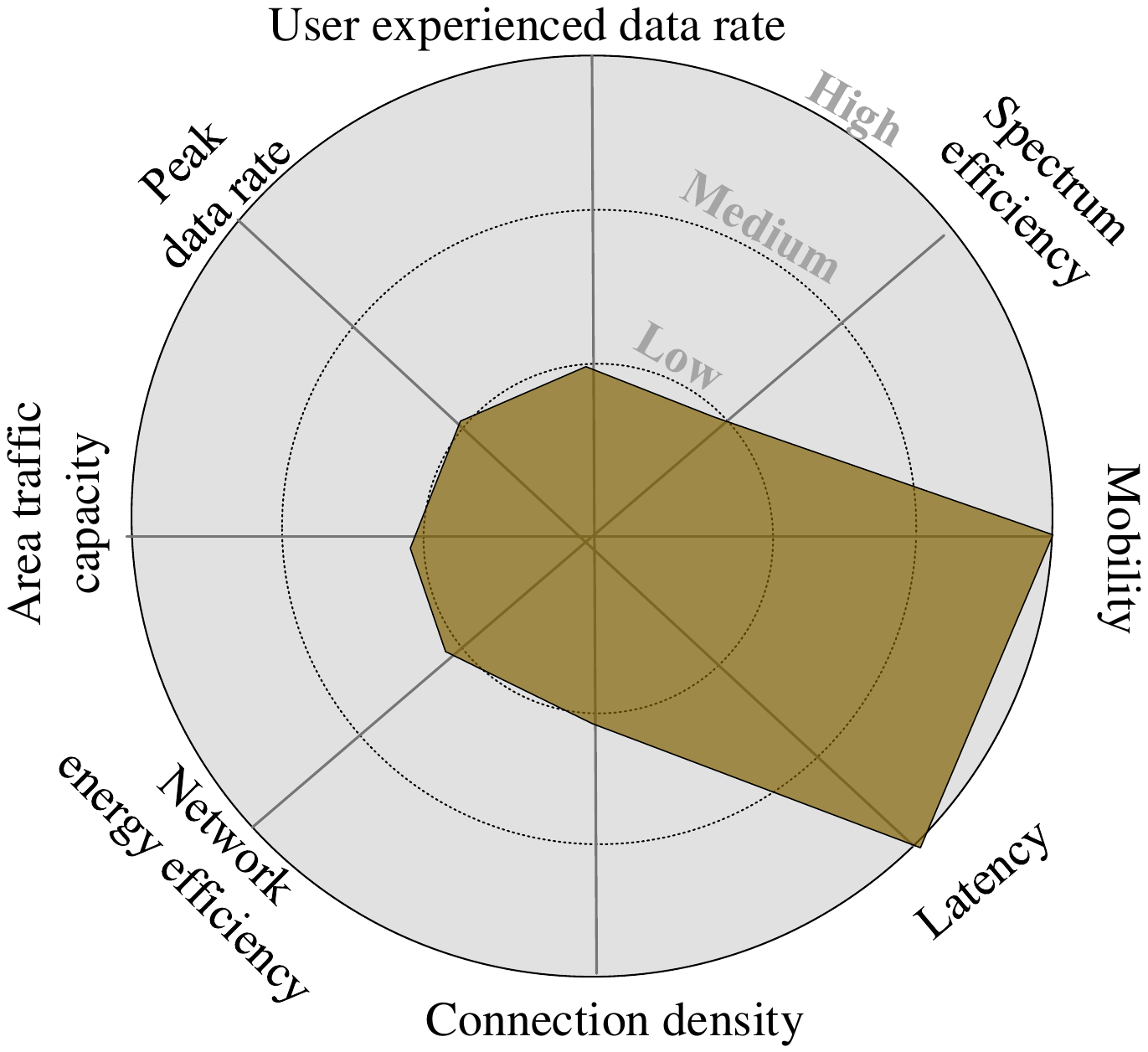}
        \caption{Critical safety applications.}
        \label{fig:safety5G}
    \end{subfigure}
     \vskip\baselineskip
     \begin{subfigure}[b]{0.46\textwidth}
        \centering
        \includegraphics[width=\textwidth]{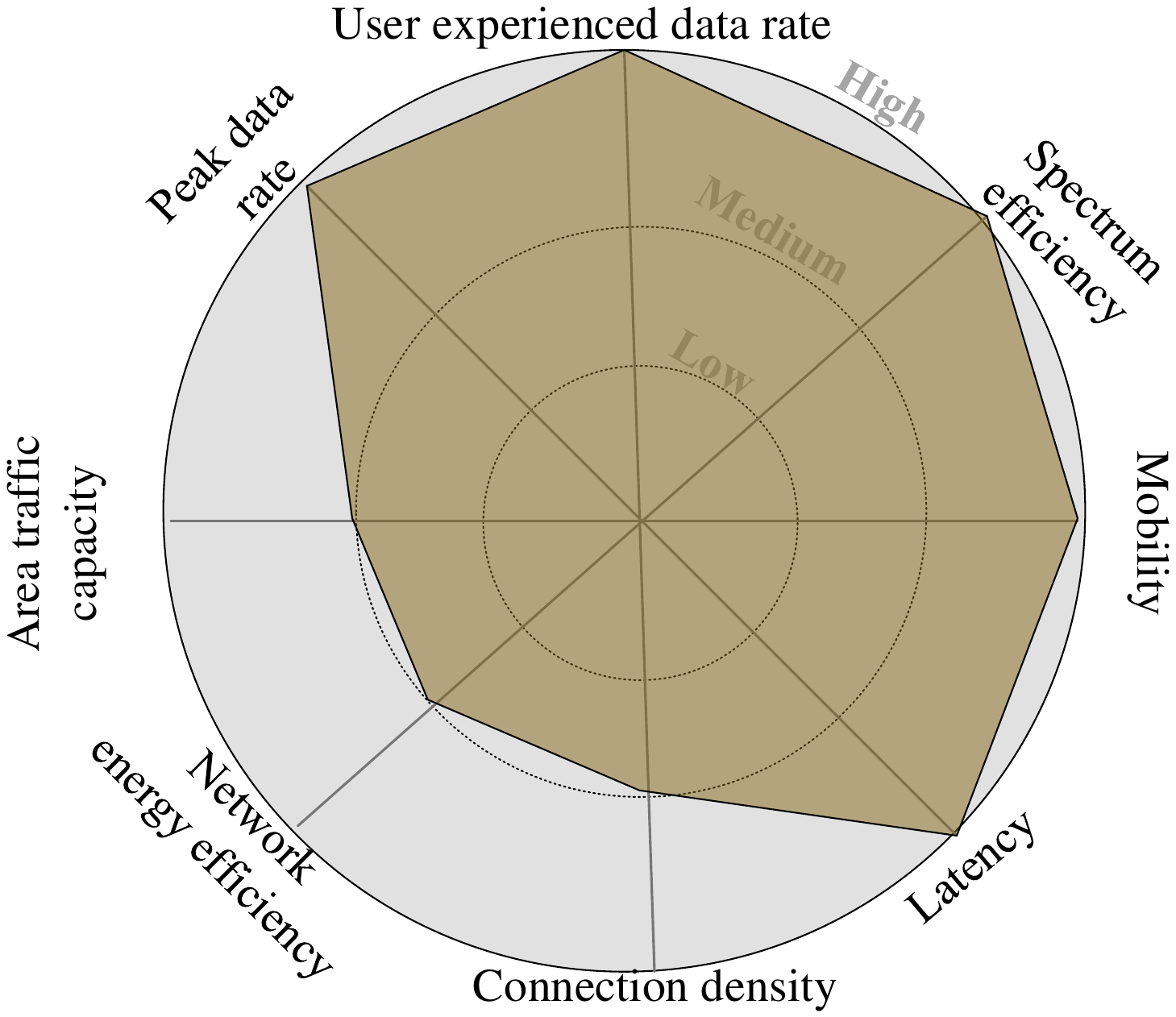}
        \caption{Autonomous driving.}
        \label{fig:ad5G}
    \end{subfigure}
    \hfill
    \begin{subfigure}[b]{0.46\textwidth}
        \centering
        \includegraphics[width=\textwidth]{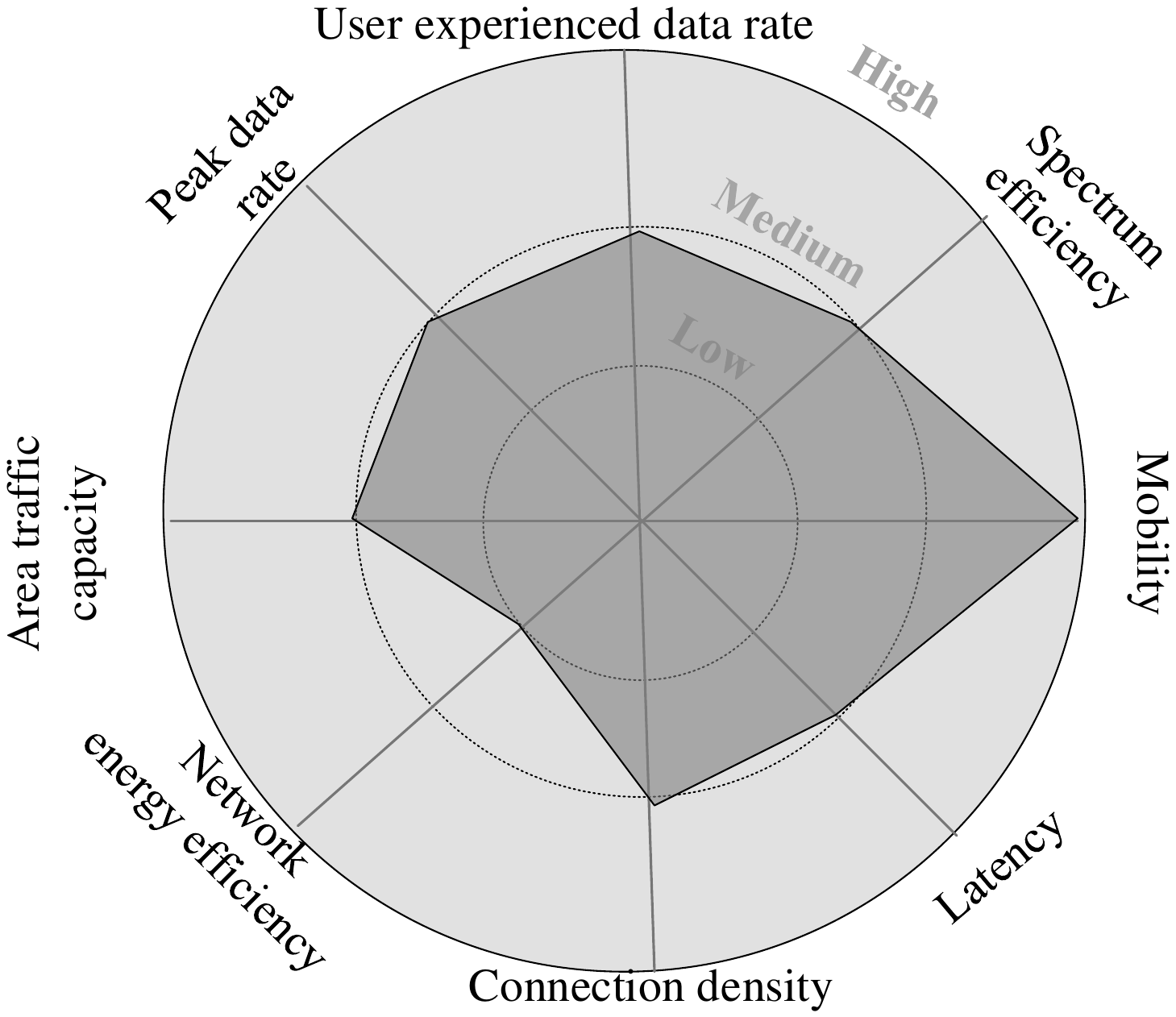}
        \caption{Efficiency and traffic management applications.}
        \label{fig:eff5G}
    \end{subfigure}
    
    \caption{The importance levels (High, Medium, Low) of key performance indicators for different classes of applications.}
    \label{fig:v5gusecases}
\end{figure*}

The international telecommunication union (ITU~\footnote{https://www.itu.int}) envisions the capabilities of future mobile networks in the international mobile telecommunications-2020 (IMT-2020) standard. The capabilities entail flexibility, reliability and security when providing various services in three intended usage scenarios, enhanced mobile broadband (eMBB), ultra-reliable and low-latency communications (uRLLC), and massive machine-type communications (mMTC). ITU sets the guidelines for 3GPP to create and maintain the technical standards for 5G technologies. 

The 5G ecosystem and defined use cases (e.g., enhanced mobile broadband (eMBB) and ultra-reliable low-latency communication (uRLLC)) are promising enablers of ITS services and applications. For instance, passengers can watch an HD movie while the driver is using augmented reality applications to detect road hazards with real-time and visually interactive navigation (usage of eMBB). Figure~\ref{fig:5gcran} illustrates the orchestration and architecture to achieve network slicing and vehicular applications' use cases. Figure~\ref{fig:v5gusecases} illustrates the importance level (Low, Medium, or High) for the key capabilities of 5G to address these use cases. Figure~\ref{fig:info5G} illustrates the importance level of the key capabilities of 5G to address the infotainment applications such as video streaming, augmented and virtual reality, and mobile cloud gaming for passengers during commuting. These applications belong to the eMBB use case, whereas critical safety and time-sensitive applications belong to the uRLLC use case, featuring stringent requirements for reliability, latency, and continuous, seamless connectivity~\cite{storck20195g}. Figure~\ref{fig:safety5G} illustrates the importance level (Low, Medium, or High) of the core capabilities of 5G to address critical safety applications. Autonomous driving (AD) requires ultra-high reliability, low latency, and high bandwidth, a combination of uRLLC and eMBB use cases. Figure~\ref{fig:ad5G} illustrates the importance level for the key capabilities of 5G to address AD requirements. Efficiency and traffic management applications are more resilient and less dependent on latency and reliability compared to safety applications. Figure~\ref{fig:eff5G} illustrates the importance level for each 5G KPI to serve these applications.

\subsection{Comparisons of performance of short-range communication standards}
There has been extensive research into comparisons between DSRC and LTE C-V2X technologies \cite{ansari2021joint}. Such comparisons have shown that the best technology depends somewhat on the deployment scenario (e.g., dense urban roads vs. highways) and the application. For example, 5G Automotive Association (5GAA) and Papathanassiou et al. have conducted extensive experiments to compare the performance of DSRC and C-V2X radio technologies for their suitability to deliver vehicle-to-everything broadcast safety messages. They confirm that LTE C-V2X significantly outperformed DSRC in various key areas~\cite{5g2019v2x,papathanassiou2017cellular}. Therefore, LTE C-V2X seems to be the promising candidate to enable these ITS services and applications. However, DSRC has undergone several large-scale field trials and is already in production in the US, Europe, and Japan. In fact, the coexistence of both DSRC and LTE C-V2X is likely in many regions. Therefore, Ansari et. al.~\cite{ansari2021joint}, for example, emphasize the need to enable V2X communication regardless of the underlying technology (DSRC or LTE C-V2X). As such, a hybrid V2X system is a potential comprehensive solution. The hybrid V2X scheme could apply spectrum sharing techniques such as frequency division multiplexing with a guard band. Additionally, performance and scalability issues of both IEEE 802.11p DSRC and LTE C-V2X/PC5 Mode 4 have been driving the future developments of IEEE 802.11bd DSRC and NR C-V2X \cite{ansari2021joint}. The coexistence of these new future standards is an open research area.

\textbf{Summary and Take-Away Message.}
NR C-V2X is a promising vehicular technology for transmitting large data volumes directly or indirectly over short distances. However, 5G-assisted long-distance communication (e.g., vehicle-to-cloud) will still incur significant latency as the data must traverse the backhaul (core) network. For instance, when a vehicle sends sensory data to a manufacturer cloud server (where the data is processed). Edge computing, thus, is a useful paradigm to enable data processing in geographic proximity to the vehicle, thus lowering the latency and enabling time-critical applications.

\section{Vehicular Computing Overview}
\label{sec:compreview}
External (to the vehicle) computing resources are important for an ITS as such resources help with the aggregation and fusion of heterogeneous data from multiple road users (thus providing a holistic view)~\cite{golestan2016situation}, and with enabling of complex applications so that these applications are accessible regardless of the processing power and storage capabilities of the vehicles. For example, traffic management, emergency management, fleet management, and Intelligent navigation (e.g., through augmented reality overlays on the windshield) are complex applications that might require relocation of complex tasks to external compute and storage resources~\cite{5g2015autovision}. These resources can be remote computing resources (such as cloud servers) or intermediary nodes such as multiaccess edge computing (MEC) servers and fog computing nodes.
% between the vehicle and the remote cloud 

%Cloud Computing
\subsection{Cloud Computing}
The NIST (American National Institute of Standards and Technology) defines cloud computing as a model for enabling ubiquitous, convenient, and on-demand access to a shared pool of configurable computing resources (e.g., networks, servers, storage, applications, and services). These resources can be rapidly provisioned and released with minimal management effort or service provider interaction~\cite{mell2011nist}. Cloud computing has so far been the dominant paradigm in terms of offloading vehicular-intensive computation. For instance, Toyota's connected car architecture is powered by Microsoft Azure HDInsight to process millions of events a day. Toyota provides its vehicles with a data communication module to transmit the vehicular data to a Toyota smart center. The latter provides a mobility services platform that enables public companies to offer Toyota and Lexus vehicles specific services. In other words, SMEs will be able to provide technological solutions which will comply with the overall system. 

Beyond traditional cloud services, the huge vehicular fleets on our roadways, streets, and parking lots can be seen as massively underutilized computational resources. Given this framing, \textbf{Vehicular Cloud Computing (VCC)} has also emerged as a new hybrid technology that incorporates vehicular ad-hoc networks and cloud computing. In this paradigm, the underutilized vehicle's resources, computing power, internet connectivity, communication resources, and storage, are shared or rented over the Internet to various customers~\cite{gu2013vehicular}. Through seamless and decentralized management of cyber-physical resources, VCC provides third-party or community services at low cost and enables efficient utilization of vehicle resources. Additionally, due to vehicle mobility, agility and autonomy, VCC can dynamically adapt its managed vehicular resources allocated to an application according to the dynamically changing requirements and system conditions. However, such a paradigm still faces high relative latency and high communication costs~\cite{ahmad2017characterizing,grover2018real}. In practice, stationary vehicles or mobile vehicles are controlled by cyber-physical resource management software to form VCCs. VCCs can thus be categorized into two classes: static VCCs and dynamic VCCs. These classes are suitable for different vehicular cloud services or applications~\cite{gu2013vehicular}. VCC viability is further enhanced by 5G deployment as 5G provides capabilities, such as large bandwidth, ultra-reliability, low-latency, and V2V communication through 5G side-links, that will support even more VCC use cases.

The proven economic benefits of cloud computing make it likely to remain a permanent feature of the future computing landscape. However, the network overhead and latency of remote cloud computing cannot meet the requirements of time-critical applications and thus proves detrimental to overall network performance. Additionally, cloud computing lacks context-awareness that, for example, captures spatio-temporal traffic and driving patterns.

 \subsection{Edge Computing}
Edge computing (EC) is a distributed computing paradigm that places computational resources and storage geographically close to end users (for example, vehicles and RSUs). Thus service requests typically travel a much shorter physical distance (and traverse fewer network nodes) for processing compared to requests to typical centralized cloud servers. This results in significantly lower latency. Additionally, EC can complement cloud computing by masking transient cloud outages and can naturally better capture contextual and situational information due to the proximity to end users~\cite{zhang2017mobile}. Overall, EC promises to deliver scalable, reliable, and low latency cloud services.

\begin{figure}[!t]
    \centering
    \includegraphics[width=1\linewidth]{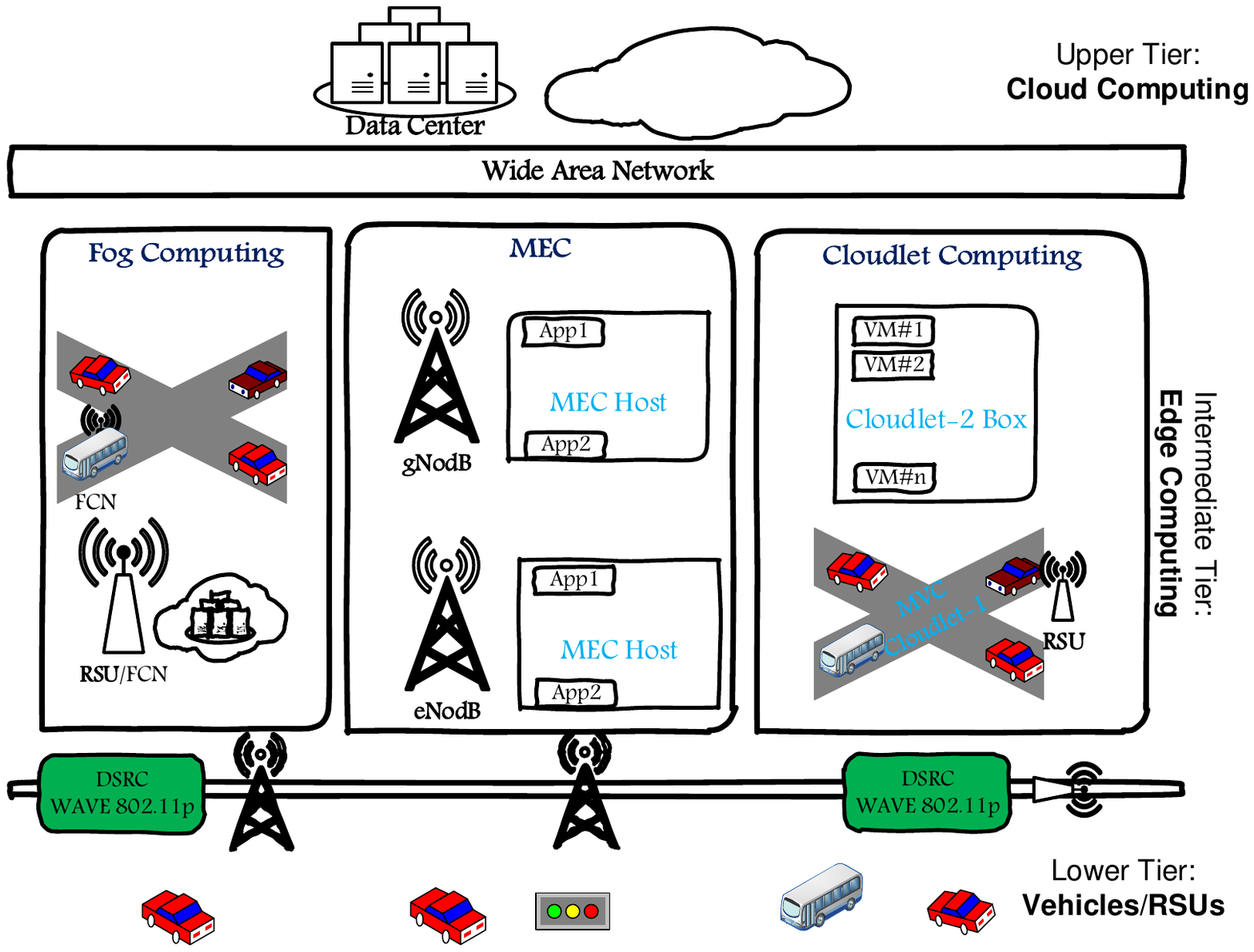}
    \caption{3-Tier architecture for an edge computing paradigm as an extension of cloud computing~\cite{dolui2017comparison,yao2015migrate}.}
    \label{fig:3tiers_edgecomp}
\end{figure}

Edge computing encompasses three distinct frameworks in the context of vehicular networks: vehicular fog computing (VFC), multiaccess edge computing (MEC), and mobile vehicular cloudlets (MVCs). Figure~\ref{fig:3tiers_edgecomp} illustrates the architectures of these three frameworks. 
 
\begin{table*}[!t]
\begin{minipage}{\textwidth}
  \begin{center}
    \caption{Summary of Cloud and Edge Computing Frameworks~\cite{borcoci2016fog,dolui2017comparison,wang2017survey}}
    \label{tab:compsummary}
    \footnotesize 
    \begin{tabular}{p{0.08\textwidth} p{0.2\textwidth} p{0.2\textwidth} p{0.2\textwidth} p{0.2\textwidth}}
      \toprule % <-- Toprule here
       \cellcolor{Gray} & \textbf{CC} / \textbf{VCC} & \cellcolor{Gray}\textbf{Fog Computing} & \textbf{MEC} & \cellcolor{Gray} \textbf{Cloudlet} \\
       \midrule % <-- Midrule here
        \cellcolor{Gray}Origin & Amazon / Olariu et al~\cite{olariu2011taking} & \cellcolor{Gray} Cisco~\cite{cisco2015fog} &  ETSI~\cite{etsi2019mec} & \cellcolor{Gray} Satyanarayanan et al~\cite{satyanarayanan2014cloudlets} \\
      \midrule % <-- Midrule here
        \cellcolor{Gray} Deployment Location & Data center / stationary and mobile vehicles~\cite{basagni2013mobile,gu2013vehicular} & \cellcolor{Gray} At any point between vehicles and cloud & Radio access network & \cellcolor{Gray} Local or outdoor installation \cite{satyanarayanan2009case,satyanarayanan2014cloudlets} \\
       
      \midrule % <-- Bottomrule here
         \cellcolor{Gray} Deployed Nodes& Dedicated servers / underused vehicle resources  &  \cellcolor{Gray} VFC (vehicles)~\cite{hou2016vehicular}, RSUs~\cite{huang2017vehicular}, or connected ESs~\cite{tordera2016fog,grover2018real}  &  ESs running in BSs aggregation or core of RAN & \cellcolor{Gray} MVC-adjacent smart vehicles and  RSUs~\cite{wang2016serviceability} Datacenter in box~\cite{satyanarayanan2009case,satyanarayanan2014cloudlets} \\

      \midrule % <-- Bottomrule here
       \cellcolor{Gray} Access Technology & Internet & \cellcolor{Gray} WiFi, mobile networks~\cite{dolui2017comparison} &  Mobile networks (LTE, 5G)  % <-- Short (DSRC, ITS-G5)  
        & \cellcolor{Gray} WiFi~\cite{dolui2017comparison},  DSRC or ITS-G5 \\
        
      \midrule % <-- Bottomrule here
       \cellcolor{Gray}  Proximity ~\cite{dolui2017comparison} & Many hops, 10s to 1000s of km & \cellcolor{Gray} One or multiple hops between vehicles and cloud & One hop, 100s of meters to few km  & \cellcolor{Gray} One hop, nearby \\

      \midrule % <-- Bottomrule here
       % &  &  &     &   \\
     \cellcolor{Gray}  Context awareness \cite{dolui2017comparison,wang2017survey} & No & \cellcolor{Gray} Medium &   High  & \cellcolor{Gray} Low \\
     
      \bottomrule % <-- Bottomrule here
       \cellcolor{Gray} Latency & High, AWS$\approx$196$\pm$84ms\footnote{Aws ping test (latency): \url{ping.varunagw.com/aws}}, Azure$\approx$176$\pm$96ms\footnote{Measure your latency to Google cloud platform(gcp): \url{www.gcping.com/}}, Google$\approx$172$\pm$106ms\footnote{Azure latency test: \url{www.azurespeed.com/}}   & \cellcolor{Gray} Low &   Low, up to 19.9 ms~\cite{pzhou2018arve}  & \cellcolor{Gray} Low, few ms\cite{satyanarayanan2009case,satyanarayanan2014cloudlets} \\

      \bottomrule % <-- Bottomrule here

    \end{tabular}
  \end{center}
\end{minipage}
\end{table*}

\textbf{Multiaccess Edge Computing (MEC)} is an edge architecture standardized by ETSI~\footnote{\url{https://www.etsi.org/technologies/multi-access-edge-computing}} that brings edge computing to the mobile network context. Specifically, MEC locates computing resources at the edge of the mobile access network, typically at the first aggregation level (base stations)~\cite{hadzic2018server}. Being an open standard, MEC also creates a standardized and open environment that enables operators to open their radio access network (RAN) edge to authorized third-parties, to flexibly and rapidly deploy innovative applications. This new ecosystem allows multi-vendor vehicles, manufacturers, and transportation agents to integrate their applications for more convenient digital services efficiently. MEC also enables applications and services to be hosted on top of the mobile network elements~\cite{etsi2019mec, etsi2019v2x}. Different deployment scenarios address various performance, costs, scalability, and operator deployment preferences:
\begin{itemize}
    \item Deployment at the radio node (eNB or gNB).
    \item Deployment at aggregation points (LTE EPC or 5GC).
    \item Deployment at the edge of the Core Network (e.g. in a distributed data center, at a gateway).
\end{itemize}
Figure~\ref{fig:3tiers_edgecomp} illustrates MEC deployment where edge servers are deployed in the cellular base stations - LTE evolved Node B or 5G NR base station (gNodeB). 

\textbf{Fog Computing (FC)} in the vehicular context refers to any intermediary computation, storage, and network services between vehicles and the cloud~\cite{dastjerdi2016fog}. Specifically, there are rich scenarios of connectivity and interactions in vehicular networks: vehicle to vehicle, vehicle to access points, smart traffic lights and roadside units (using Wi-Fi, DSRC), vehicle to network (using LTE, 5G), and other V2X scenarios. For instance, a smart traffic light node interacts locally with many sensors, which detect the presence of pedestrians and bikers, and measure the distance and speed of approaching vehicles. The smart traffic light in this context acts as a fog computing node (FCN). A fog computing node (FCN) can be any node with communication, computation, and storage resources. As shown in Figure~\ref{fig:3tiers_edgecomp}, a FCN can be a moving or parked vehicle, also referred to as vehicular fog computing (VFC)~\cite{birhanie2018mdp}, a roadside unit, or an edge device installed in a cellular base station. \textbf{Vehicular Fog Computing (VFC)} aggregates the abundant resources of individual and connected vehicles and exploits their available computing resources to enhance the application quality of service. VFC uses moving and parked vehicles as FCNs to offload computation tasks and provide networking services~\cite{hou2016vehicular,xiao2019quantitative,ning2019vehicular}. The fog has several characteristics which make it the ideal platform and non-trivial extension of the cloud to deliver services in infotainment, safety, traffic efficiency, and analytics. These characteristics are 1) low latency, 2) wide-spread and geo-distributed deployment, 3) location, mobility, and context-awareness, 4) interoperability, federation, and heterogeneity (deployable in various environments), and 5) support for real-time interactions~\cite{klas2015fog,bonomi2012fog}.

\textbf{Cloudlet Computing} represents an architecture with auxiliary proximate cloud resources for providing highly responsive services. Specifically, these cloud resources can be viewed as delegates or proxies of the real cloud and are located at the middle tier of a three-tier hierarchy, as shown in Figure~\ref{fig:3tiers_edgecomp}. A cloudlet can be either a mini data center in a box~\cite{grover2018vehicular,park2018advanced}, or vehicular resources referred to as a \textbf{mobile vehicular cloudlet (MVC)}. As an example use case, during cloud or backhaul outages, the cloudlet takes over the responsibilities and masks the outage~\cite{satyanarayanan2013role}. Adjacent vehicles and roadside units can connect via DSRC communication or 5G sidelink to form MVCs. Thus MVCs harness the computational resources of the adjacent nodes in a timely and efficient manner via peer-to-peer communication. An MVC is a cluster of smart vehicles and RSUs located in a region. Such Vehicles and RSUs can share resources and information via V2V communication or indirectly via V2I communication~\cite{koukoumidis2011pocket,gu2013vehicular}.

\noindent Overall, edge computing architectures have similarities such as the purpose. However, they have slight differences in the origin, the deployment location, the involved nodes, the access technologies, the geographical proximity, the level of contextual awareness, and the latency. Table~\ref{tab:compsummary} summarizes the differences between edge computing architectures (FC, MEC, Cloudlet) and compares them with cloud computing architectures (CC, VCC).

\textbf{Summary and Take-Away Message.}
The synergy of NR C-V2X and edge computing (EC) drops the end-to-end latency significantly, allowing stakeholders to use bandwidth efficiently, and enabling time-critical applications to run in real-time. Edge computing reduces the networking overhead and provides highly responsive services for the users on wheels. Moving the computation to the edge pushes the utilization efficiency of the next-generation mobile network to its limit. VCC increases the utilization efficiency by using the dispersed underutilized resources of the vehicles. Vehicular networks can be employed to remotely offload latency-tolerant computation (into moving or parking vehicles) and storage services (into parking vehicles), or locally offload latency-sensitive computation (into moving vehicles) and caching (into moving vehicles).

\section{Data Analytics: Technologies and Methodologies Integration.}

\label{sec:datanaltics}
\begin{figure}[!t]
     \centering
    \includegraphics[width=0.48\textwidth,keepaspectratio]{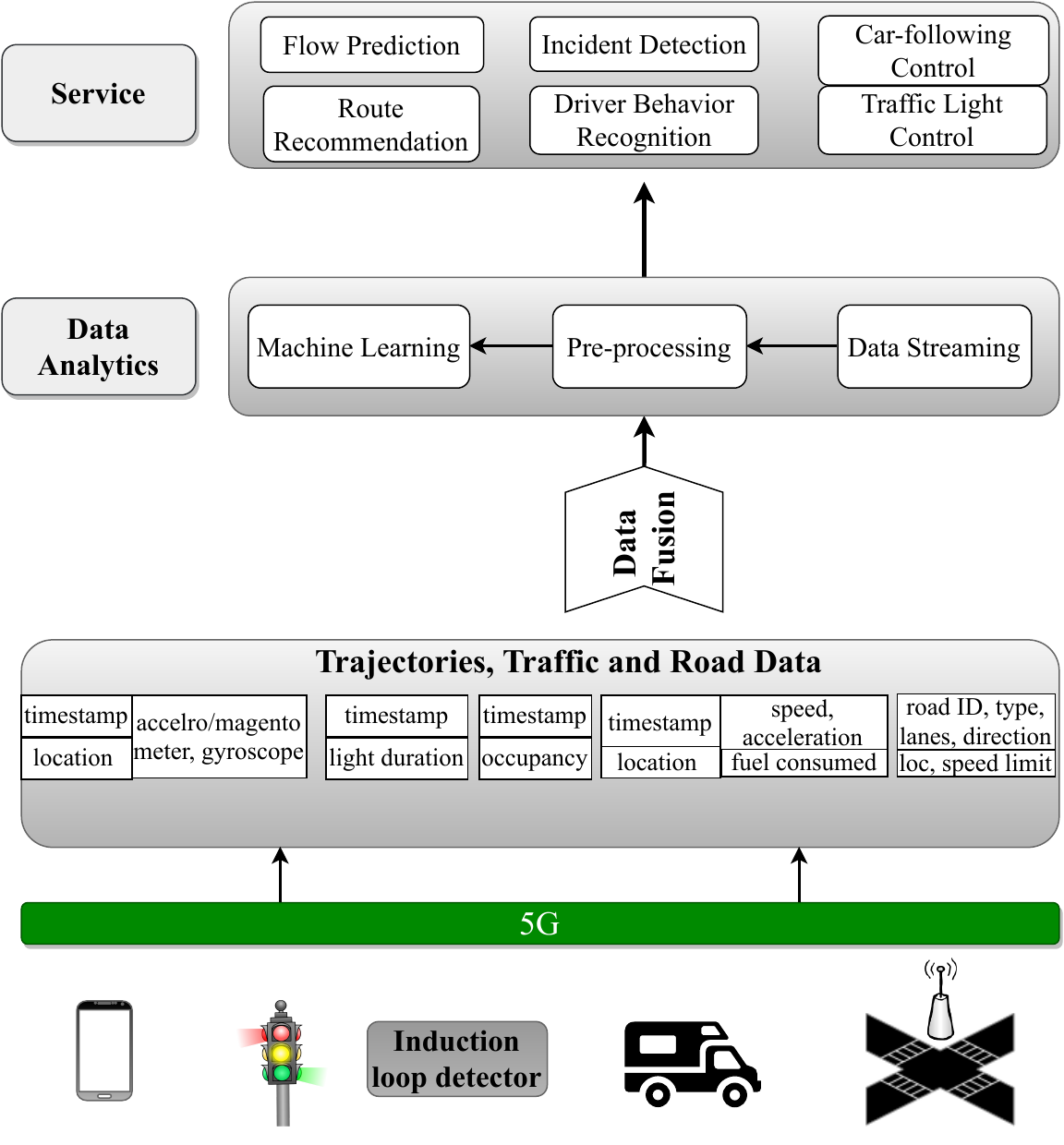}
    \caption{A general data analytics pipeline illustrating traffic data collection, aggregation and fusion, streaming into micro-batches, and processing.}
    \label{fig:pipeline}
\end{figure}

Road users and elements (e.g., Pedestrians' smartphones, vehicles, lampposts, traffic lights, or other RSUs) generate both mobility and service-related data which is heterogeneous and large in volume. Analysing this data and extracting useful and relevant information in real-time requires an efficient data analytics architecture. This architecture must support multiple data sources, allow large data volumes, enable data streaming (to achieve low latency), and allow developers to plug-in queries and machine learning algorithms. A variety of specific technologies and frameworks can be combined to actually realize such an architecture, we briefly describe some of the most common technologies and frameworks. Additionally, we study several specific architectures from literature and two case studies.

\subsection{Data Analytics Technologies}
Firstly, in terms of data storage technologies, Hadoop Distributed File System (HDFS) is a distributed file system for reliably storing large amounts of unstructured, semi-structured, and structured data as files (typically on disk). HDFS was one of the first large-scale distributed file systems for big data. Several other big data storage systems actually build on top of HDFS. HBase, for example, is a key-value pair NoSQL database with master-slave replication that leverages HDFS as underlying storage. Other notable systems include Cassandra, a popular key-value pair NoSQL database with asynchronous masterless replication.

In terms of data messaging, collection, and aggregation, the current dominant system is Apache Kafka\footnote{\url{https://kafka.apache.org/}}. Apache Kafka is a distributed event streaming system. Specifically, Kafka provides a distributed publish-subscribe messaging system that allows for decoupling of different stages of data pipelines. Kafka accommodates big heterogeneous data and Kafka event streaming includes true (event at a time) streaming with exactly-once semantics.

Finally, in terms of actual distributed computing and ML, Hadoop MapReduce is a distributed computing framework for the parallel processing of large datasets often stored on disk on HDFS (though other storage solutions are also supported). MapReduce runs on a Hadoop cluster and often leverages a cluster manager like YARN to schedule applications and services on the cluster and manage the cluster resources like memory and CPU. In comparison to the primarily disk-based MapReduce, Apache Spark~\footnote{\url{https://spark.apache.org/}} is a unified big data analytics engine for distributed in-memory data processing. Furthermore, Spark provides both batch and stream processing, libraries for Machine Learning, and an SQL-like interface. Relatedly, Apache Flink is also a big data analytics engine for distributed processing. A few of the major differences between Spark and Flink are that Spark is more mature with a larger community, while Flink was designed specifically for stream processing and thus provides better support for true (event at a time) streaming. In contrast, Spark primarily supports micro-batch streaming. Kafka also provides some true (event at a time) stream processing functionality (through the Kafka streams API).

\subsection{Integration of Technologies towards Vehicular Computing} 

\begin{figure}[!t]
     \centering
    \includegraphics[width=0.45\textwidth,keepaspectratio]{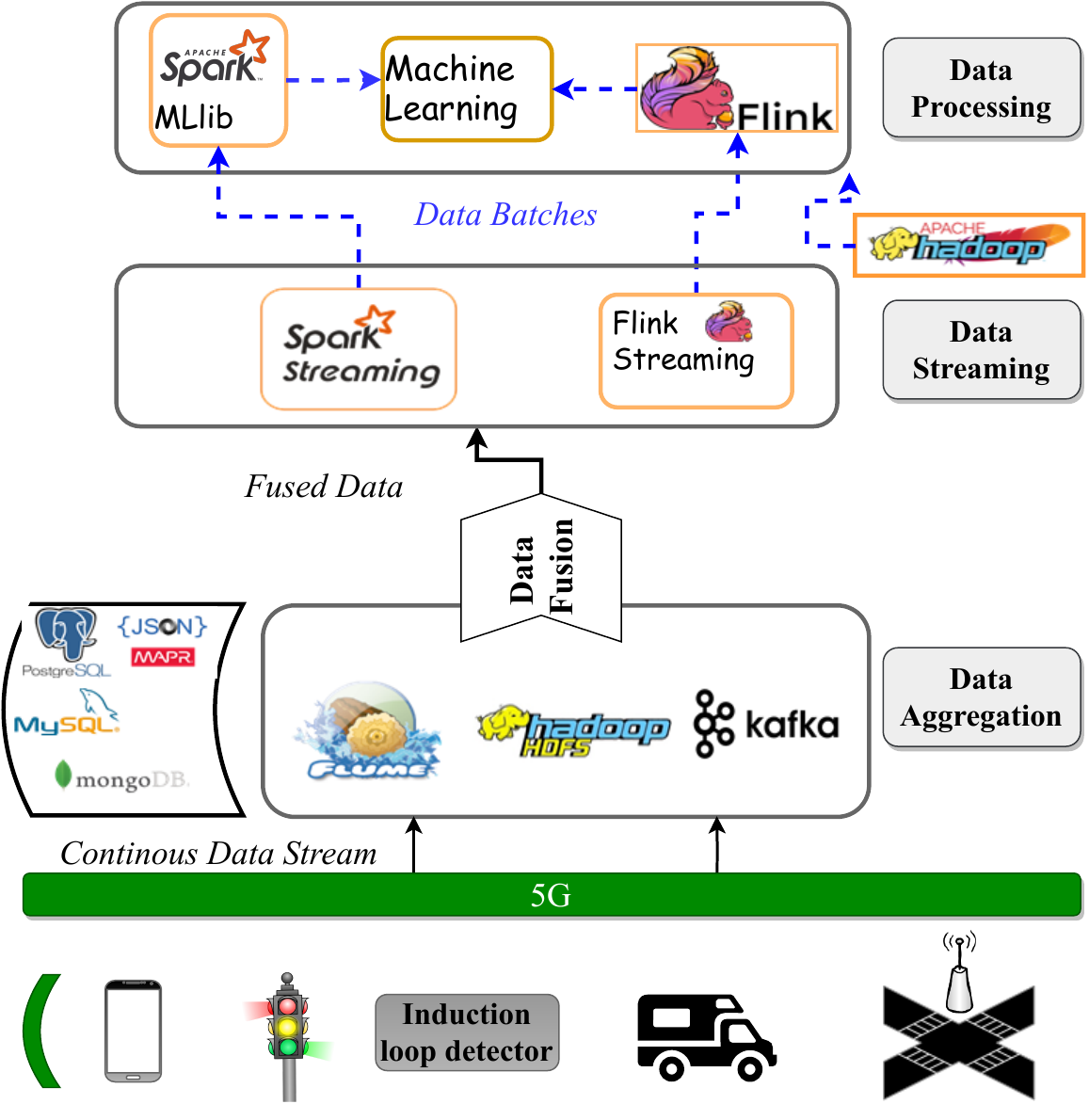}
    \caption{Example open source technologies for a data analytic pipeline. The stages include online and historic traffic data collection, aggregation, and fusion (using Kafka, Hadoop, or Flume), micro-batch streaming (using Flink or Spark Streaming), and processing including machine learning (using Flink or Spark MLLib).}
    \label{fig:techintegation}
\end{figure}
 
In vehicular environments, the big data analytics architecture relies on the integration of such technologies into a pipeline that enables computation offloading. 

Figure~\ref{fig:techintegation} illustrates a potential pipeline that integrates various technologies to process the traffic data in real-time. Firstly, Apache Kafka ingests the live vehicular/traffic data and partitions the data into distinct topics which enables multiple readers and writers to operate simultaneously thus improving scalability. A separate data fusion module facilitates fusing and aggregating the topics' data for richer features and better context determination. Spark Streaming then either consumes data from specific Kafka topics or retrieves data from HDFS, and splits the data into micro-batches to feed into SparKMLlib which applies ML algorithms. Apache Flink could replace spark to achieve a very similar setup. The ML algorithms output meaningful information (such as predictions) which is often sent back to applications or services to assist driving or traffic control, or stored permanently in stable storage such as Hadoop HDFS or Cassandra for later use. For instance, Amini et al~\cite{amini2017big} employ Apache Kafka to stream traffic data in real-time and control traffic lights in a distributed manner. Microsoft added Apache Kafka on Azure HDInsight to run a robust, real-time, big data streaming pipeline at enterprise scale, natively integrated Kafka with Azure managed disks, and made it globally available. Since then, large companies have been using this service in production to process millions of events per second and petabytes of data per day to power scenarios like Toyota's connected car, Office 365's clickstream analytics, fraud detection for large banks, or log analytics~\cite{kafkaDec2017azure,kafka2017azure}. 

\subsection{ML-Empowered Applications}
Traffic-related learning tasks can be primarily sorted into two main classes, \textbf{basic safety} and \textbf{advanced efficiency}. Collision warning and traffic incident detection are examples of basic safety, whereas traffic flow prediction, car-following, and driving behavior recognition are examples of advanced efficiency. To implement these tasks a machine learning algorithm is applied to traffic data in a data analytics pipeline (see Figure~\ref{fig:pipeline}). We briefly discuss the details of these different task classes including example systems from research studies.

\textbf{Traffic Flow Prediction} is the real-time short-term prediction of traffic on the road network that assists in understanding the future traffic state. This prediction can leverage both longer-term historical traffic data (e.g., diurnal patterns) and up-to-date signals of traffic conditions. Such prediction plays a significant role in road network traffic planning and traffic control optimization and lays the foundation for, travel guidance, navigation, and other mobility services. A variety of temporal ML algorithms including NN-based and more traditional algorithms have been applied to the task. The most sophisticated NN-based models including long short-term memory (LSTM), stacked LSTM, temporal LSTM, and spatial-temporal autoencoder LSTM (SpAE-LSTM) outperform the more traditional multilayer perceptron (MLP) model, decision tree model, and support vector machine (SVM) models for traffic flow (congestion) prediction ~\cite{chen2016long,mou2019tlstm,lin2019spatial}. In terms of details, SpAE-LSTM, for example, is a hybrid model consisting of a sparse autoencoder and an LSTM. The sparse autoencoder captures the spatial features while the LSTM captures the temporal features \cite{lin2019spatial}. In fact, many of the related models leverage autoencoders as they can learn generic traffic flow features and obtain the internal relationship of traffic flow~\cite{liu2018urban, wei2019autoencoder,lv2014traffic}. 

In addition to short-term prediction, trend-modelling of the traffic time series can facilitate longer-term traffic forecasting. Such forecasting relies on the implicit temporal correlations among the time series observed on different days/locations due to human diurnal patterns. Specifically, the daily traffic time series at a certain location have similar M-shapes over consecutive days, in which the morning and evening rush hours correspond to the two peaks of the M-shape~\cite{li2015trend}. Li et al~\cite{li2015trend} use principle component analysis (PCA),  a well-known mathematical dimensionality reduction and feature extraction procedure, to project the traffic time series onto an n-dimensional orthogonal linear space such that the data with the $k^{th}$ largest variance by projection lies on the $k^{th}$ dimension. PCA trend can establish a link between traffic time series collected in different days/locations because the observed daily data series share the same set of latent variables. PCA also can assist to predict whether the traffic is normal or abnormal by comparing the distances between their projections in the latent space.

\textbf{Traffic Incident Detection} is the detection of real-world traffic incidents in some given spatiotemporal area. This detection is essentially a mining task from heterogeneous traffic data. An interesting data source for incident detection is social media such as Twitter and Facebook (which are popular and real-time in nature). Specifically, the public often immediately posts and shares information when traffic incidents occur (e.g., road closures, traffic congestion, or accidents), which spreads rapidly through the network. The user messages shared in social networks are called status update messages (SUMs). These messages may contain text and meta-information such as timestamp, geographic coordinates (latitude and longitude), name of the user, links to other resources, hashtags, and mentions. Several SUMs referring to traffic incidents in a limited geographic area may provide valuable information about abnormal traffic incidents. 

Text mining and natural language processing (NLP) are the specific ML areas relevant for the extraction of useful information and knowledge from unstructured text (e.g., Tweets and Facebook posts)~\cite{salas2017incident,allahyari2017brief}. Thus pre-processing and algorithms from these areas become important components in the data analytics pipeline for such a detection application. In terms of research implementations, D'Andrea et al.~\cite{d2015real} present a real-time traffic event monitoring system that leverages Twitter stream analysis. They employ SVM to classify tweets as traffic event related with very high accuracy (>95\%). Relatedly, Traffic Events Detection and Summary (TEDS)~\cite{liu2014search} uses NLP, spatial-temporal mining, and wavelet analysis techniques to create a traffic incident map with text summarizations (from multiple same-incident tweets) from Twitter data.

\begin{table*}[!t]
  \begin{center}
    \caption{Summary of Machine Learning Methods for ITS Applications}
    \label{tab:aisummary}
    \footnotesize 
    \begin{tabular}{p{0.2\textwidth} p{0.2\textwidth} p{0.2\textwidth} p{0.2\textwidth} p{0.12\textwidth}}
      \toprule % <-- Toprule here
      \textbf{ITS Application} & \textbf{Phenomenon} & \textbf{ML method} & \textbf{Data Source}& \textbf{Deployment}\\
     
      \midrule % <-- Midrule here
        Distraction Detection &  Using mobile phone while driving  & Deep Learning -CNN~\cite{celaya2019texting} & Ceiling-mounted camera &  Vehicle OBUs  \\
        \cmidrule(rl){3-4}
        &    & \cellcolor{Gray} CNN and RNN~\cite{streiffer2017darnet}  &  \cellcolor{Gray} Inward facing camera and IMU data &  \\
        
      \bottomrule % <-- Bottomrule here
        Traffic flow prediction &   Congestion & Stacked LSTM~\cite{chen2016long}, Temporal LSTM~\cite{mou2019tlstm}, Spatial-Temporal,  LSTM~\cite{lin2019spatial} & Traffic Data & RSU \\\\
        \rowcolor{Gray}
     Traffic flow, speed prediction &  Traffic dynamics & CNN, RNN, SAE and autoencoder~\cite{liu2018urban}  & Traffic Data from Infrastructures, Trajectory AFC Records \& Social media    & - \\
    
      \bottomrule % <-- Bottomrule here
      Traffic Accident Detection   & Traffic anomalies  & SVM~\cite{d2015real}  & Social networks & - \\\\
      \bottomrule % <-- Bottomrule here
        
    Driving Detection \& iDentification, $D^3$& Abnormal driving    & SVM~\cite{chen2015d}  & Smartphone sensors  & Smartphone app \\\\
    \cmidrule(rl){1-4}
    \cellcolor{Gray} Drowsy Driving Detection ($D^3$)  & \cellcolor{Gray} Drowsy driving & \cellcolor{Gray} LSTM ~\cite{xie2019d}  & \cellcolor{Gray} Embedded Acoustic Sensors in Smartphones   &   \\

      \bottomrule % <-- Bottomrule here
      Cooperative Lane Changing   & Traffic Competition  & Deep RL~\cite{wang2019cooperative}  & Vehicle on-board sensors & Vehicle OBUs \\\\
      
      \bottomrule % <-- Bottomrule here
      \rowcolor{Gray}
      Safe, Efficient and Comfort Following  & Car Following  & MDP, Deep RL~\cite{zhu2019safe}  & Vehicle on-board sensors & Vehicle OBUs \\

      \bottomrule % <-- Bottomrule here
      
      Intelligent Cross-Layer\& Cooperative Offloading & Diverse requirements, Time-varying of Content Popularity & Deep Deterministic Policy Gradient (DDPG)~\cite{dai2019artificial}  & Vehicle's OBUs, RSUs, Environment & RSUs, Base stations  \\\\
      \bottomrule % <-- Bottomrule here

    \end{tabular}
  \end{center}
\end{table*}
 
\textbf{Vehicle following} is the systematic control of vehicles' velocity to optimize and maintain safe, comfortable, and convenient traffic flow. To this end, car-following models determine the velocity of a following vehicle in response to actions of a lead vehicle~\cite{zhu2018modeling}. Many works have developed reinforcement learning (RL) models which control the vehicle's velocity to optimize the traffic flow. Meixin et al~\cite{zhu2019safe} develop a deep RL model which uses a reward function reflecting driving safety, efficiency, and comfort to fulfil the multi-objectives of the car-following model. They design the RL reward function to combine driving features and maximise the cumulative rewards and which leads to more efficient traffic flow compared to human drivers and faster running speeds compared to model prediction control. Collision avoidance strategy is incorporated for safety and faster convergence. 

Relatedly, Wang et al.~\cite{wang2019cooperative} use deep RL to control lane-changing behavior for each vehicle with a reward function defined as a trade-off between the vehicle's travelling efficiency (i.e., how efficiently a vehicle maintains a target speed), traffic flow rate, and level of cooperation between the vehicles. Specifically, they utilise a deep neural network named deep Q-network (DQN) as the RL model and the lane-changing of each vehicle is formulated as a Markov decision process (MDP). The state space $S(t)$ is the vehicle's state (at a given time $t$) which consists of three sequential frames of traffic snapshots and the corresponding speed difference between actual and target speed. The action space $A(t)$ is the corresponding driving decision such as switch to left or right lane, speed up by a fixed increment up to a maximum speed, or maintain the current speed. Walraven et al~\cite{walraven2016traffic} also apply MDP, Q-learning, and neural networks to learn policies dictating the maximum allowed driving speed on highways to reduce traffic congestion. The above-mentioned systems are based on RL with a reward function that improves the overall traffic efficiency instead of the travel efficiency of an individual vehicle. In short, cooperation leads to a more harmonic and efficient traffic system rather than competition.

\textbf{Driver Behavior Recognition} is the classification of a driver’s behavior into classes such as normal, aggressive, or distracted driving. The classifier output is conveyed to the drivers audibly, visually, or haptically using an infotainment system to raise the driver's awareness and allow them to react in time. Raising the driver's awareness improves their driving behaviors and thus promotes safer driving, reduces traffic accidents, and contributes to social safety~\cite{wu2016driving}. Distracted driving is an especially relevant and dangerous driving behavior. Distracted driving can be defined as any activity that diverts the driver's attention from driving including talking or texting on a mobile phone~\cite{li2018texting} or using on-board entertainment or navigation systems. In the US, 3,166 people in 2017~\cite{nhtsa2018distracted} and 3,477 people in 2015 died in motor vehicle crashes involving distracted drivers.

Celaya-Padilla et al~\cite{celaya2019texting} leverage a ceiling-mounted wide-angle camera that feeds data to a convolutional neural network (CNN) to detect distracted drivers. The detection of distracted driving can then be conveyed to the driver audibly, visually, or haptically using, for example, the infotainment system. Relatedly, DarNet~\cite{streiffer2017darnet} is a framework utilizing CNNs and recurrent neural networks (RNNs) to process images (from a driver-facing camera) and inertial measurement unit (IMU) data (from the driver's mobile device) to detect distracted driving behavior. These data sources provide rich contextual information that allows fine-grained. For instance, an image of a driver sending a text message can be cross-validated by checking the acceleration of the mobile device from the embedded accelerometer. Such multi-modal cross-validation improves the classification accuracy without the need to deploy additional sensors. More generally, Driving behaviour Detection and iDentification system ($D^3$)~\cite{chen2015d} also detects abnormal driving behaviors using real-time smartphone sensors and an SVM-based ML algorithm. These driving behaviours include, for example, weaving, swerving, sideslipping, fast u-turn, turning with a wide radius, and sudden braking.

In addition to distracted driving, drowsy driving is another problematic behaviour that threatens road safety. Sober-Drive system~\cite{xu2014sober} is a smartphone-assisted drowsy driving detection system that uses the smartphone's front camera feed and analyses the open/closed states of the driver's eyes using a NN model. Thus the system leverages drowsiness indicators such as the eyelid closure percentage, blink time, and blink rate. Furthermore, the D3-Guard system~\cite{xie2019d} detects drowsy driving using audio recording by smartphones and a long short term memory (LSTM) network. The system detects nodding, yawning, and abnormal steering in real-time by leveraging the Doppler shift of the audio signals to capture the unique patterns of these drowsy driving actions. In these systems, model training typically occurs offline whereas the application uses the real-time smartphone sensory data for inference in an online phase.

\textbf{Summary and Take-Away Message.}
ITS applications often require accuracy and real-time while handling big and heterogeneous data. Therefore, the underlying platform must perform the entire pipeline in real-time, including data ingestion, streaming, processing, plugging in the ML model, and output presentation. Leveraging and combining distributed data analytics technologies such as Kafka and Spark can fulfill these requirements. Though the specific setup and placement of such pipelines will vary significantly given the diversity in requirements and scope of different ITS applications.

\section{Integrated Communication and Computing Architecture}
\label{sec:arch}

Some vehicular applications require offloading computation tasks to external servers. In this section, we study potential architectures that support such applications under three different scales: local scale (i.e., road, intersection, or last-mile), neighborhood scale, and city scale.
\subsection{Local Scale}
\label{sucsec:micro}

%Figure local scale
\begin{figure}[!t]
    \centering
    \includegraphics[width=.5\textwidth,keepaspectratio]{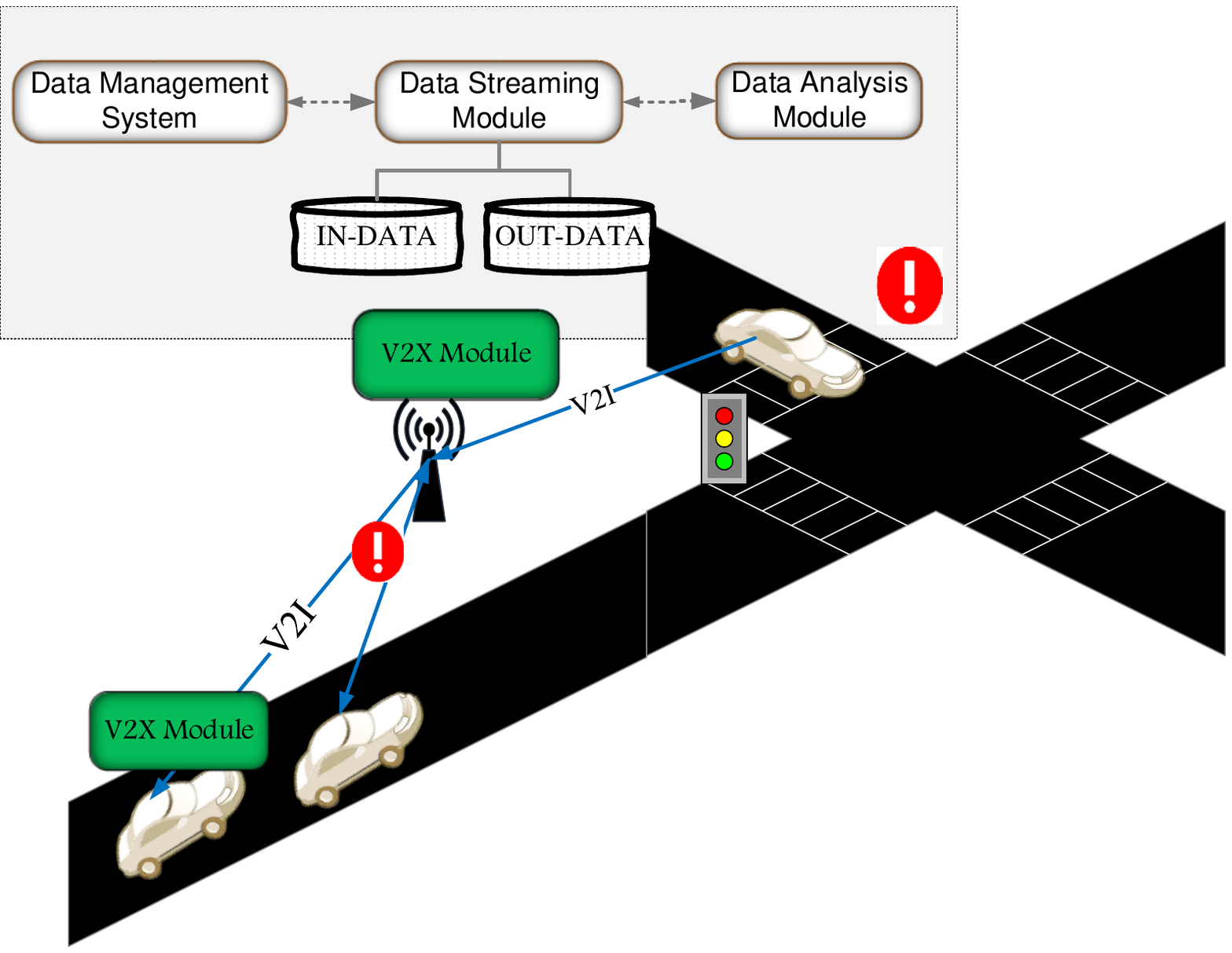}
    \caption{Integrated V2X and RSU-assisted computing architecture at local scale}
    \label{fig:localscale}
\end{figure}

\begin{figure*}[!t]
    \centering
    \begin{subfigure}[b]{0.58\textwidth}
        \centering
        \includegraphics[width=\linewidth]{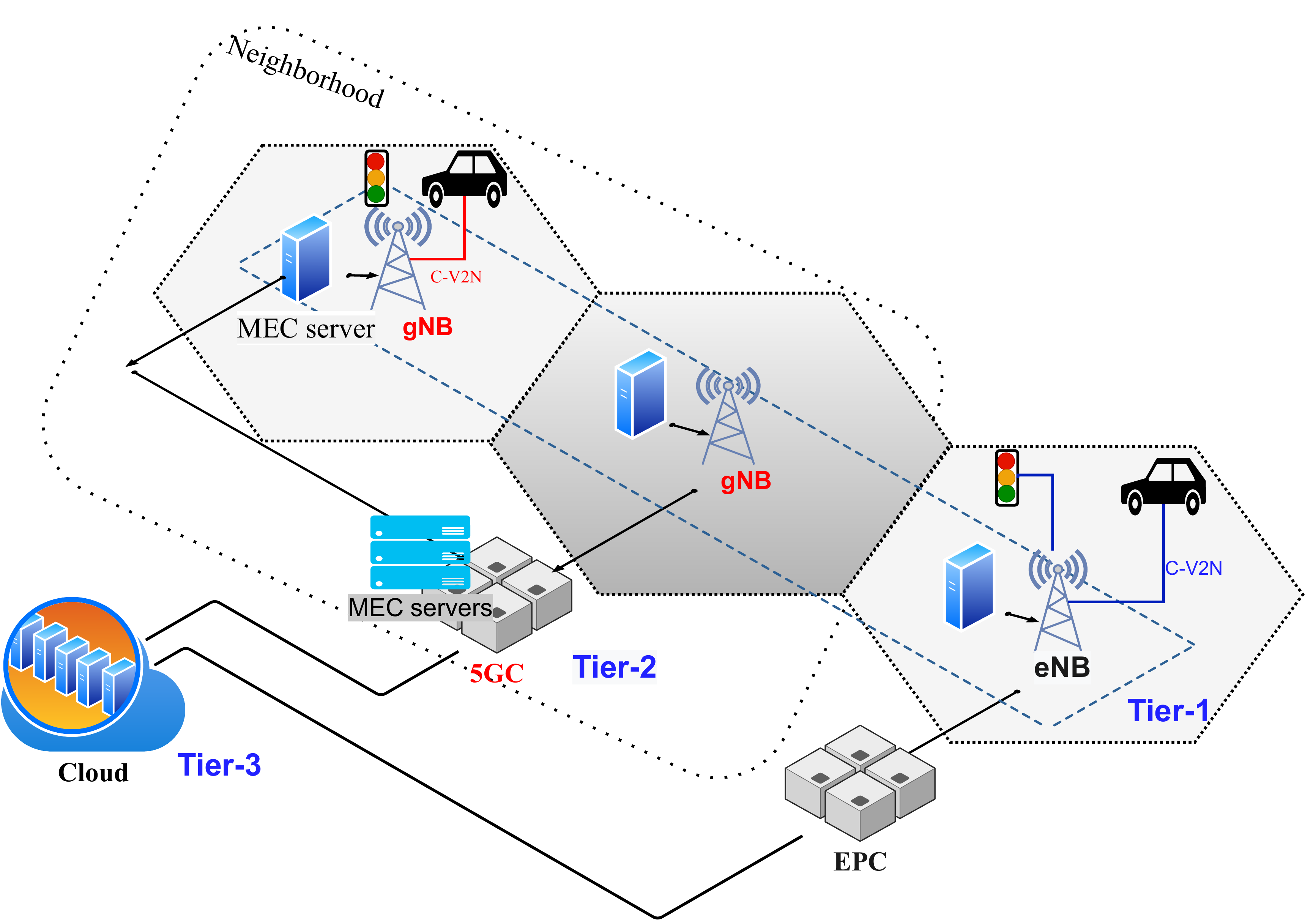}
        \caption{Communication-aware neighborhood and city scale examples}
        \label{fig:comm_neighborhood}
    \end{subfigure}
    \hfill
    \begin{subfigure}[b]{0.4\textwidth}
        \centering
        \includegraphics[width=\textwidth,keepaspectratio]{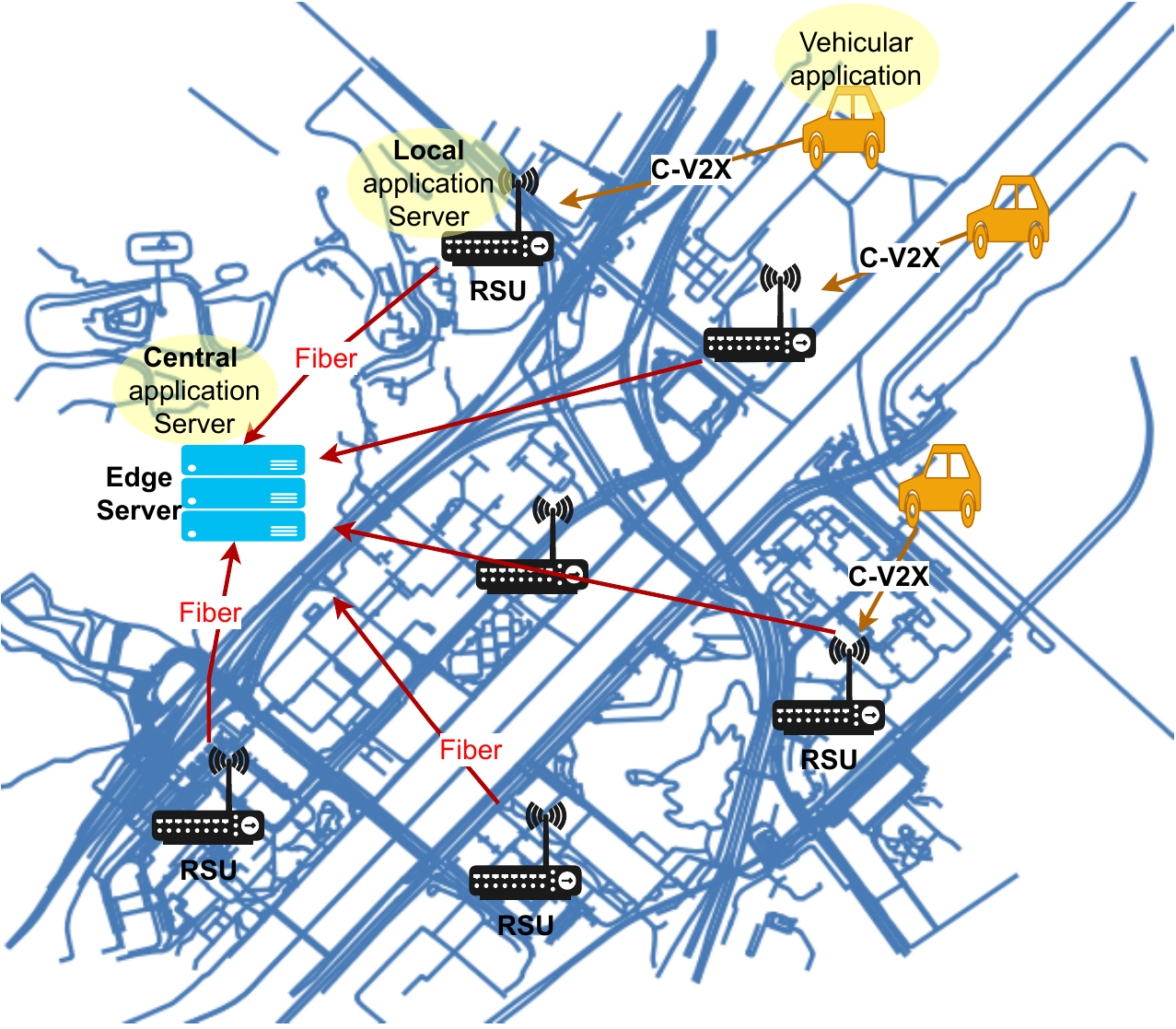}
        \caption{Transport-aware neighborhood scale example. }
        \label{fig:trans_neighborhood}
    \end{subfigure}
    
    \caption{Scale from communication and transport perspectives. (a) Multiple base stations (local-scale areas) form a neighborhood scale. Tier-2 denotes neighborhood scale, whereas Tier-3 denotes city scale. Vehicles and traffic lights communicate with base stations (gNB and eNB) using C-V2N.  (b) RSU range forms a local-scale area, and multiple RSUs form a neighborhood scale and collaborate through edge servers. The infrastructure is overlaid on the road network (blue lines) of the Shatin neighbourhood of Hong Kong.}
    \label{fig:neighborhoodscale}
\end{figure*}

At the local level (road/intersection), the first priority is to provide \textbf{basic safety} functions to prevent accidents. For instance, active safety applications warn drivers of impending danger so the driver can take corrective or evasive action. Beyond basic safety, \textbf{advanced efficiency} functions also play a local level role. As an example, an active traffic management system (an advanced efficiency function) may adapt the local traffic control system proactively or reactively to improve local travel flow. Such a system includes consideration of peak-hour traffic, the detection of and response to incidents, and the reduction of waiting time due to congestion and incidents. Thus the system enhances the local transportation network performance in terms of safety, efficiency, reliability, scalability, and sustainability~\cite{Xiong2021atm}. 

\textbf{Requirements.}
Decisions at this scale have to be made very quickly thus stressing the importance of low overall system latency (normally less than 100ms).

Basic road safety requires ultra-reliable and low latency communication. The infrastructure must have the capability to monitor the traffic situation reliably and make accurate decisions. For example, the road-side sensors must be capable of accurately recognizing and localizing various types of objects (e.g., vehicle, pedestrian, obstacle) with low latency. In advanced efficiency, the system must observe the situation and ambient environment and take quick actions on local scales to ensure smooth traffic flow. For instance, the system might change the duration of a traffic light phase based on road occupancy. Though the latency requirement of advanced efficiency system is on the order of seconds as such delays only minimally impact performance.

Importantly for both basic safety and advanced efficiency, local driving patterns vary according to the spatio-temporal context. For instance, mean driving speed varies fairly predictably due to time of day (rush hours, night hours, and so forth), day of the week (weekday or weekend), and road type (motorway or street road). Moreover, traffic flows change due to abnormal traffic events such as road incidents. Therefore, many applications on the local scale require context-awareness as well as situational-awareness to make accurate decisions~\cite{alhilalcad3, satyanarayanan2013role}. Additionally, the vehicles must maintain continuous, uninterrupted, and highly available communication between each other and with the RSUs. Finally, dynamic vehicle mobility leads to rapid topology changes in VANETs; while a variety of events cause transient unbalanced traffic distributions and congestion on roads and intersections. Therefore, such a category of applications should scale well with traffic flow and the incurred channel bandwidth usage~\cite{hartenstein2010vanet}. 

\textbf{Architecture and Methodology:} To ensure context-awareness (in advanced efficiency) and low-latency (in basic safety), a local-scale system should offload computing tasks to nodes close to road users. Figure~\ref{fig:localscale} illustrates a detailed example system that includes these architectural considerations and meets the aforementioned local requirements. Computing nodes co-locate alongside roads (RSUs), or with cellular base stations (LTE-eNB and 5G-gNB). Relatedly, offloading computation to edge nodes provides distributed and parallel computing, allowing the system not only to scale up with load but also to ensure higher reliability by avoiding congestion in the back-haul network. Additionally, C-V2X allows vehicles to communicate with RSUs, thus allowing the collection and distribution of additional vehicular data (beyond the data from an RSUs own sensors). 

The architecture contains three essential components: a data streaming module, a data management system, and a data analysis module to cope with the continuous traffic data streams. Traffic data includes but is not limited to the vehicle data, the environment data collected from the area's installed sensors, and the information obtained from the RSU. The data streaming module ingests and aggregates the traffic data streams then fuses both the aggregated data and potentially stored data from the data management system. Afterwards, the streaming module divides the continuous data flow into batches for processing in the data analysis module~\cite{peppes2021driving,alhilalcad3,amini2017big}. For real-time streaming, existing works~\cite{peppes2021driving,alhilalcad3} integrate Apache Kafka and Apache Spark to detect unsafe driving activities. To provide real-time streaming and analytics of traffic data, they integrate Apache Kafka, to ingest the continuous stream of traffic data, and Spark Streaming to divide the stream into periodic micro-batches. The micro-batches then feed a Spark engine in which they are plugged into a machine learning algorithm to analyse the data. Amini et al~\cite{amini2017big} propose a similar architecture but rather use a reducer-evaluator instead of Spark to analyse the data. The architecture is distributed and flexible architecture to control traffic signals in real-time. It employs Apache Kafka as an intermediary module between the traffic system with its sensors ( e.g., probe vehicles, loop detectors) and actuators (e.g., traffic lights, variable message sign), and the data analysis module. As data come in, they are being processed via user-specified reducer functions. After a time interval, a separate evaluator function is invoked to assess the results and update the settings of the traffic system accordingly.

\subsection{Neighborhood Scale}
\label{subsec:meso}

Another category of (vehicular) applications involves decision-making on a larger neighborhood scale. From a transportation perspective, the neighborhood scale encompasses multiple interconnected (via road network) local-scale areas with a fairly reasonable traffic flow. From a mobile network perspective, it typically encompasses multiple adjacent macrocells (see Figure~\ref{fig:neighborhoodscale}).

%Requirements
\textbf{Requirements:} Traffic flow (inbound and outbound) and the underlying road network must drive the decision-making at neighborhood scale. Decisions are made based on traffic interactions between local areas and their cascading effects, and the acceptable latency to take action. Such context-awareness allows the system to disseminate accident warnings to the relevant road users, balance road occupancy, alleviate congestion, and thus reduce travel time and CO2 emissions. Importantly, the latency requirements at this scale are typically less stringent (e.g., on the order of seconds or minutes) than the local scale. Yang et al~\cite{yang2019mobilityaware} consider the task's maximum latency (threshold) while minimizing the required communication and computing resources to implement a mobility-aware task offloading scheme. The scheme balances task completion latency and necessary computation and communication resources.

\textbf{Architecture and Methodology:}
To ensure a holistic neighborhood view, a neighborhood scale architecture often co-locates the computational units with the aggregation points of a mobile network, known as communication-aware neighborhood scale, or the aggregation points of the RSU system, known as transport-aware neighborhood scale, as shown in Figure~\ref{fig:neighborhoodscale}. For example, Zhou et al~\cite{zhou2019erl,pzhou2021drle} optimize neighborhood-scale decision making through communication-aware MEC servers on two different tiers, collocated with base stations and aggregation points. In addition to the three local scale architecture components, the neighborhood architecture often enables a collaboration channel between computing nodes to exchange information (processed data), sometimes through a central computing and storage node. Some systems~\cite{alhilalcad3,yang2019mobilityaware} enable a neighborhood or larger view through inter-RSU collaboration whereas Zhou et al~\cite{zhou2019erl,pzhou2021drle} use two-tier MEC computing as shown in Figure~\ref{fig:comm_neighborhood}.

CAD3 integrates data streaming and processing components for real-time decisions and in-time reactions~\cite{alhilalcad3}. At the core of data analytics, multi-agent reinforcement learning (MARL) and federated learning (FL) can be utilized to capture the changing traffic patterns and maintain a smooth traffic flow~\cite{zhou2019erl,pzhou2021drle}. However, dual-mode C-V2X roadside device, supported by mobility-aware algorithms would allow the road users to intercommunicate via NR C-V2X/PC5 direct communication channel (5.9GHz band) and allow them to connect to network infrastructure via the Uu (5G) communication channel~\cite{cv2x2018astri,cv2x2018hkt}. Dual-mode communications allow the road users within areas on local scale to share vehicular movement and traffic status information with others, leading to a coverage of neighborhood scale (see Figure~\ref{fig:cityscale}).

\subsection{City or Larger Scale}
\label{subsec:macro}
The last category of applications require city-scale decision making such as advanced efficiency (e.g., city traffic management and planning), holiday out-of-city traffic inference, and studying of big events (i.e., sporting and exhibition events)~\cite{zhang2019decomposition,zhang2020urbananalytic}. These often involve the \emph{think globally, act locally} concept, where data analytics of traffic big data collected from road users and elements (e.g., vehicles, traffic lights, pedestrians) covers the entire city.

 \begin{figure}[!t]
    \centering
    \includegraphics[width=0.98\linewidth]{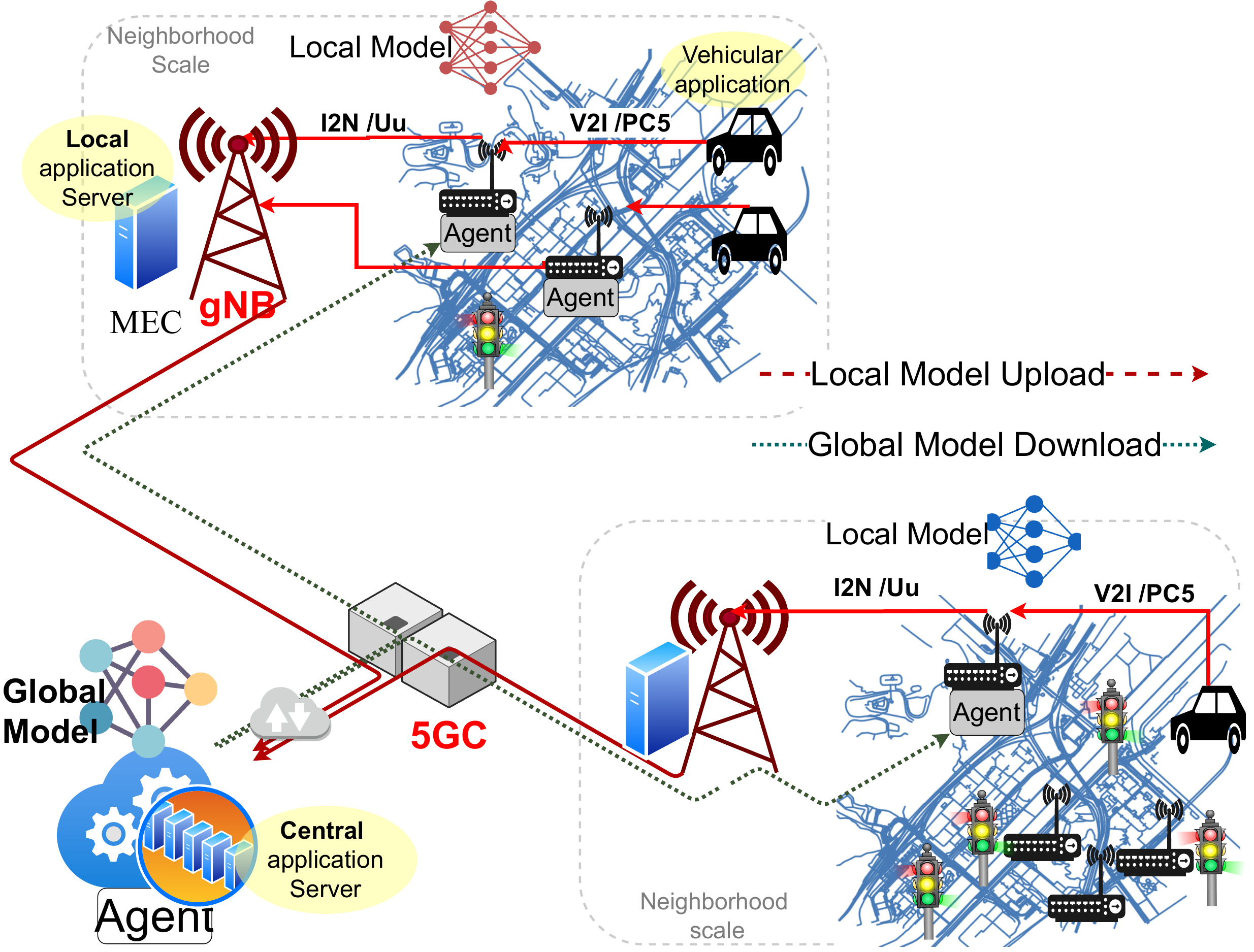}
    \caption{Combined transport and communication-aware architecture for city-scale deployment. RSU are equipped with dual-mode connectivity to receive road user data via V2I direct sidelink (PC5) communication. RSUs then forward data to the cloud via I2N Uu communication through the network infrastructure. Road users can also upload local and download global models (FL) or use local/central agent (RL) instead of sharing raw data.  }
    \label{fig:cityscale} 
\end{figure}

\textbf{Requirements:} Many city-scale applications require urban big data that is naturally high volume, high variety, and high velocity (3Vs). This data can encompass trip and trajectory data, surveillance video, weather, social events, and a diversity of traffic data. The data must be associated with time and location stamps, in other words, spatio-temporal data, to enable many rich spatio-temporal applications including, for example, finding dynamic dependencies among different regions at the local and neighborhood-scale~\cite{chawla2012inferring}.

For city traffic planning and management, transport decision-making involves understanding city-scale human mobility patterns and discovering traffic problems (traffic anomaly~\cite{zhang2019decomposition} and accident detection and congestion prediction) ~\cite{chawla2012inferring}. As such, the data analytics engine often resides in the cloud to handle the heavy computations. 

Event prediction and detection often leverages data streaming including preprocessing and feeding into a machine learning model to detect events or predict the occurrence of events~\cite{zhang2017urbandynamic,zheng2008understandgps,gonzalez2008understanding}. For instance, \textbf{traffic flow prediction} requires city-scale traffic data analytics and urban dynamics decomposition~\cite{zhang2019decomposition} to identify the traffic flow patterns and predict the future traffic flow which then helps to ensure efficient route planning and mobility. While \textbf{traffic management} requires either a multi-tier communication-aware architecture~\cite{liu2017high,zhou2019erl,pzhou2021drle}, or centralized cloud architecture for collecting and processing the massive traffic data for monitoring traffic density, throughput, and events in real time. Though communication-aware systems lack spatial-temporal correlations and mobility-awareness. Efficient traffic flow requires mobility-aware traffic control so as to decrease the waiting time of vehicles traveling on signalized roads. 

\textbf{Architecture and Methodology:} A city-level architecture primarily locates the computing nodes at the cloud given such centers' capacity to process large streams of city traffic data. In such an architecture the road elements (e.g., traffic lights, lampposts, vehicles, inductive loops) can be equipped with 5G modules to transmit the large data volumes to the nearby RSUs via NR C-V2X direct communications (PC5). The RSUs then transmit data to the cloud through the network infrastructure, i.e., I2N Uu communications. The cloud ingests the continuous stream of widespread urban sensor data and uses data streaming and analytics engines (e.g., Apache Kafka and Spark) to process them in real-time including machine learning algorithms for decision making. Road elements receive data from the cloud to enable vehicular applications. For instance, traffic lights receive commands via downlink Uu communications, decision-makers (department of transportation) also receive information via downlink Uu communications, other road users (e.g., vehicles and pedestrians) receive directions via downlink (Uu) and then I2V/PC5 communications (see Figure~\ref{fig:cityscale}). For traffic management, distributed agents can monitor and take actions at the local scale and neighborhood scale while a city-scale central agent tunes performance parameters to optimize the overall traffic flow. However, the transmission of raw traffic data may still cause issues due to high bandwidth requirements and privacy restrictions (for example related to General Data Protection Regulation (GDPR))~\cite{voigt2017eu}. A potential methodology for dealing with these issues is federated learning (FL). FL addresses privacy and bandwidth issues by training local models and uploading only the local model parameters to the cloud for aggregation into a global model (that is then redistributed to the local agents)~\cite{du2020federated} (see Figure~\ref{fig:cityscale}).

\textbf{Summary and Take-Away Message.}
A robust vehicular computing architecture is crucial for satisfying the requirements of vehicular applications. These applications vary with some requiring detailed local spatio-temporal context with low latency decision making, and others requiring a city-wide holistic view for optimization beyond the local or neighborhood scale. Some applications will even require some combination of these requirements. Incorporating the V2X and 5G ecosystem, including edge computing, can help cope with such requirements. Specifically, MEC and NR C-V2X communication are likely very important. NR C-V2X is a crucial component for reliably sharing the big traffic data. While MEC enables processing the data even at the network edge thus allowing a broader range of potential latency including real-time or near real-time streaming data analytics (for example with Apache Kafka or Spark).

Overall, we note the boundary between these scales is not well defined and any traffic system or  application must still, in some ways, consider the holistic traffic environment on multiple physical scales. In any case, most systems and applications will actually target multiple physical scales.

\section{Real World Case Studies}
\label{sec:casestudy}

In this section, we study three real-world scenarios: local-scale cooperative perception, neighborhood-scale accident warning, and city-scale event detection (urban planning). These scenarios highlight the leveraged technologies, the enabling architecture, and the interactions with road users.

\begin{figure*}[!t]
    \centering
    \begin{subfigure}[b]{0.7\textwidth}
        \centering
        \includegraphics[width=\textwidth,keepaspectratio]{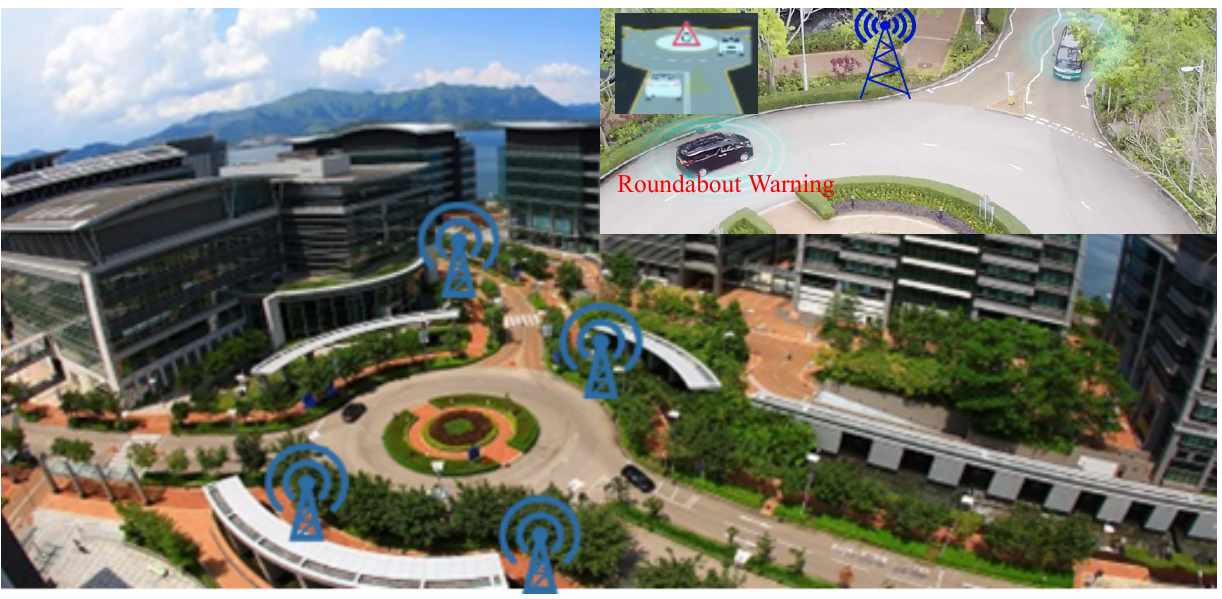}
        \caption{Local-scale roundabout warning of danger ahead using V2X and RSU~\cite{astri2021roundabout}}
        \label{fig:roundabout}
    \end{subfigure}
    \hfill
    \begin{subfigure}[b]{0.45\textwidth}
         \centering
        \includegraphics[width=\linewidth]{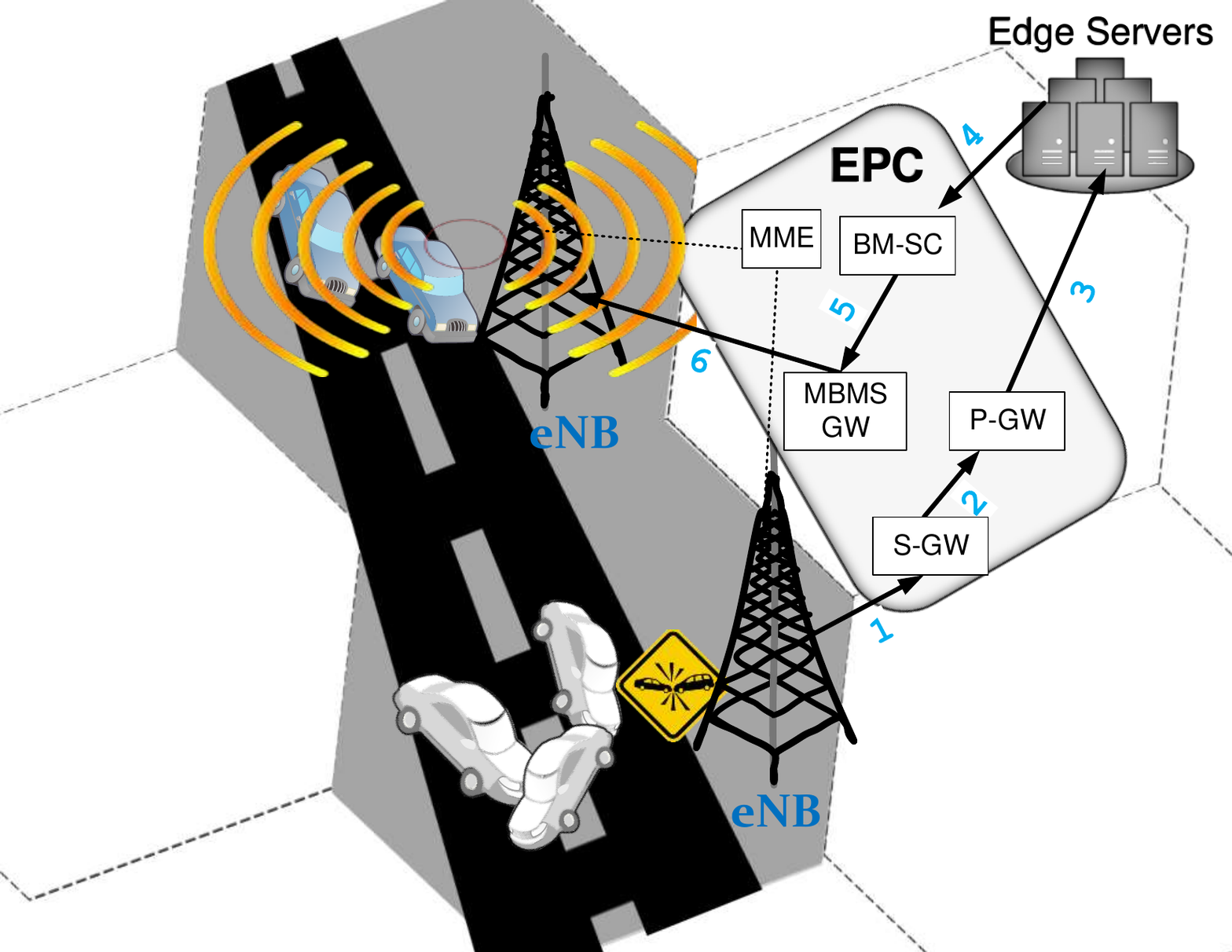}
        \caption{Broadcast delivery of accident warning using LTE eMBMS.}
    \label{fig:lteUnicastBcast}
    \end{subfigure}
    \hfill
    \begin{subfigure}[b]{0.5\textwidth}
        \centering
        \includegraphics[width=\linewidth]{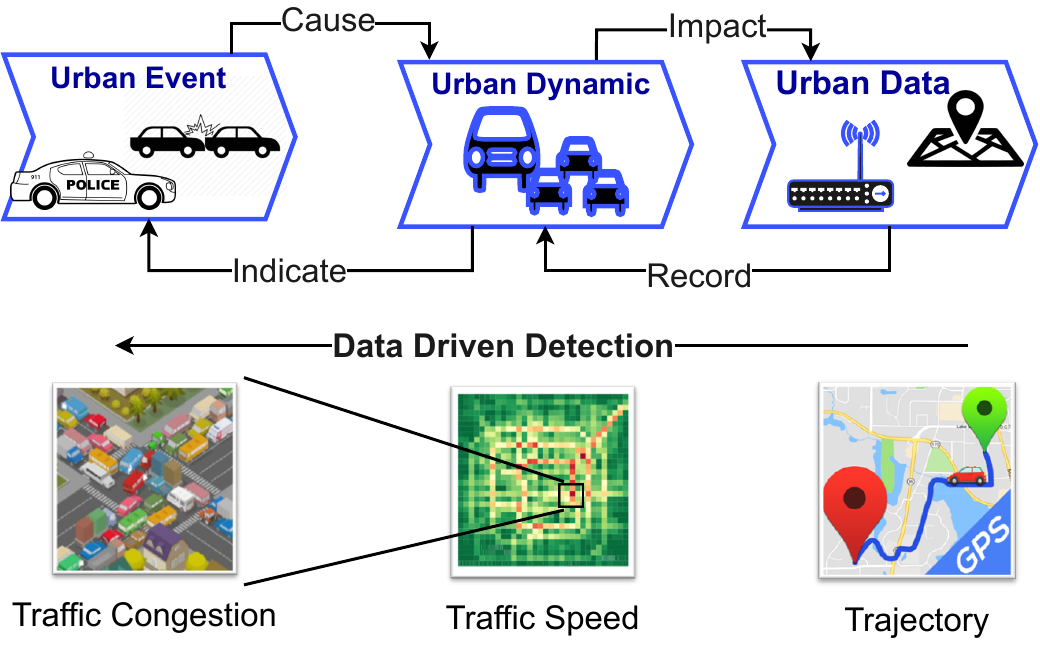}
        \caption{City-scale data-driven event (congestion) detection ~\cite{zhang2017urbandynamic}}
        \label{fig:urban.planning}
    \end{subfigure}
    
    \caption{Real-world case studies at local, neighborhood, and city scales.}
    \label{fig:realworld}
\end{figure*}

\subsection{Case study: Cooperative Perception}
\label{subsec:copercept}

Modern vehicles have a variety of sensors to perceive the nearby environment and warn the driver of potential hazards. However, external objects (e.g., buildings, other vehicles, trees) may block the view of these sensors, causing blind spots and raising road safety concerns. Additionally, each type of sensor also has inherited limitations in terms of sensing distance, accuracy, and environmental dependency. 

Figure~\ref{fig:roundabout} illustrates a classic basic safety example. Vehicles pass through a roundabout, inevitably encountering a blind spot situation that causes a safety issue. Specifically, when a new vehicle enters the roundabout, the black vehicle's driver can not see the gray oncoming vehicle, stopped vehicles, or pedestrians crossing the road behind the stopped vehicle. The black vehicle might not have an in-vehicle communication and computation module (OBU) and thus cannot identify and relay information about the stopped vehicle and crossing pedestrian. The assumption that all vehicles are equipped with OBUs does not hold in the real world. As such, the roadside system becomes essential to recognize the object type (vehicle or pedestrian), position, and speed,  accordingly determine the potential danger and notify the driver. See-around-the-corner vehicular applications allow the RSU installed at the intersection to distribute real-time sensor data or notifications to vehicles in range using NR C-V2I communications and extend the vehicles' visibility beyond the sensors and driver's visibility~\cite{sakaguchi2020towards,hakeem20205g}. Owing to the short distance and direct communication between the vehicle and RSU, the roadside system would be able to recognise the danger and disseminate corresponding warnings within 100 ms (maximum acceptable latency). 

When vehicles are equipped with OBUs, they can cooperatively perceive a larger area of the environment than could any single vehicle alone. Using C-V2X, the adjacent vehicles can communicate directly (PC5) to share raw onboard sensor data (e.g., camera, LiDAR, radar) and processed information (e.g., information about identified objects). Thus they can obtain rich dynamic information in complex traffic environments with blocked views~\cite{fukatsu2021automated}. NR C-V2X enables sharing large volumes of sensory data with a peak data rate of 1 Gbps or more due to the wide bandwidth in the mmWave region~\cite{hakeem20205g}. An example of the benefits of OBU-enabled cooperative perception is seen in vehicle overtaking situations. Specifically, in such a situation a see-through vehicular application could allow the ego vehicle to obtain the front camera view (using NR C-V2V communication) of the leading vehicle to help identify the non-line-of-sight (NLOS) traffic situation ahead ~\cite{sakaguchi2020towards}.

\subsection{Case study: Accident Warning}
Figure~\ref{fig:lteUnicastBcast} illustrates a neighborhood-scale accident warning system that leverages the EPC of an LTE network. A vehicle involved in an accident or a nearby vehicle that observed the accident sends a notification to the local evolved NodeB (eNB). The eNB delivers the notification data packet via the radio bearer to the serving gateway (S-GW), which, in turn, forwards the notification to the packet data network (PDN) gateway (P-GW). The P-GW provides an entry point for service providers (i.e., dedicated servers to collect information or notifications and disseminate them to the subscribed group). The broadcast multicast service center (BM-SC), part of the core network, functions as the interface between the distribution service (MBMS) and service provider (on edge servers), thus supporting evolved multimedia broadcast multicast services (eMBMS). The BM-SC transmits the notification as broadcast or multicast content through the eMBMS gateway (MBMS-GW) to the eNBs using IP multicast and then to the subscribed vehicles in each eNB cell. 

In addition to the traditional unicast transmission, the eMBMS can broadcast to all the users, or multicast to a predetermined set of users (drivers, passengers, or other users) in a cell using a single eNB~\cite{eMBMS3GPP}, or in adjacent cells using multiple eNBs, i.e., a multimedia broadcast multicast service single frequency network (MBSFN). Alternatively, the roadside system could also recognize the accident and forward warnings to the RSUs in affected local areas. Those RSUs could then disseminate the warnings to vehicles directly through C-V2X (I2V) or indirectly through LTE, depending on the vehicle communication module.

\subsection{Case study: Urban Planning}
Smart devices, roadside sensors, and various kinds of road users collect traffic data city-wide in real-time and form a large-scale, cross-domain and multi-view data ecosystem. Such large-scale urban data enables data-driven intelligence to detect, analyze and predict large urban events. The early detection or prediction of such large events (e.g., sporting and entertainment events, protests, weather, or natural phenomenon) allows governments to take timely actions. 

Figure~\ref{fig:urban.planning} illustrates the connection between the physical space (urban event) and the cyberspace (urban data) and data-driven detection of congestion. Urban events are a causal factor in urban dynamics which reflects in urban data. Likewise, urban dynamics can be inferred from spatio-temporal urban data and urban dynamics reveal the underlying urban events. The collection of urban spatio-temporal trajectory data in real-time can help calculate the city-wise traffic flow and localize local areas of congestion~\cite{zhang2017urbandynamic}. The detection of congestion allows the intelligent transport system (ITS) to adjust the phases of traffic light signals dynamically and change routes on users' route recommendation applications (vehicular application in Figure~\ref{fig:cityscale}) to alleviate the congestion. For instance, an ITS might arrange more taxis to the area near a soccer match and recommend unrelated vehicles alternative routes that bypass the area.

\section{Conclusion and Future Directions}
\label{sec:con}

Intelligent transportation systems are set to become a major part of future human mobility. A crucial part of ITS is vehicular networks which are spatially distributed and communicate directly through short-range or sidelink communications or indirectly through network infrastructure or roadside units. These road users (e.g., vehicles) share information and offload computation tasks to nodes close to the vehicles (i.e., edge computing), or remote (i.e., vehicular cloud computing and cloud computing).

In this work, we surveyed the existing literature on distributed vehicular communication and computation. Specifically, we highlighted several vehicular network applications (e.g., basic safety and advanced efficiency ) including their technical requirements from different viewpoints. We then detailed the enabling technologies and promising methods and architecture to support these applications. In detail, we described briefly the available communication technologies including DSRC, ITS-G5, and LTE, and comprehensively 5G and NR C-V2X. Next, we reviewed the computation and data analytics frameworks applicable to a vehicular network context including general architectures such as cloud and edge computing and more specific approaches such as vehicular fog computing and mobile vehicular cloudlets.

Vehicular communications technologies, including V2N/Uu communication to 5G network infrastructure, and V2X/PC5 direct (sidelink) communications, are maturing and will jointly support a plethora of vehicular contexts (such as varying transmission distance, latency, and vehicle density). Leveraging these communication technologies, edge computing and vehicular cloud computing combined with data analytics and streaming technologies (e.g., Apache Spark and Apache Kafka) will enable applications with varying latency windows, ranging from highly latency-sensitive (basic safety) to latency-tolerant (urban event prediction/detection).

This work provided a broad survey of vehicular communication and computing. However, many interesting topics in this area remain active research targets including, for example, 1) developing programmable communication links (such as SDN) that enable on-demand bandwidth and eliminate network bottlenecks; 2) integrating security at the node, domain, end-to-end, and service levels; and using privacy-preserving machine learning methods such as federated learning; 3) developing and applying ML and AI methods that adapt to dynamic and changing driving patterns; 4) using augmented reality to improve the quality of experience of ITS, and 5) assessing the interplay between vehicular communications and future 6G technologies and standards. We briefly describe each of these aforementioned active research targets.

\subsection{Multi-carrier Selection and Aggregation for Latency and Congestion Issues}

\begin{figure}[!hbp]
    \centering
    \includegraphics[width=0.5\textwidth,keepaspectratio]{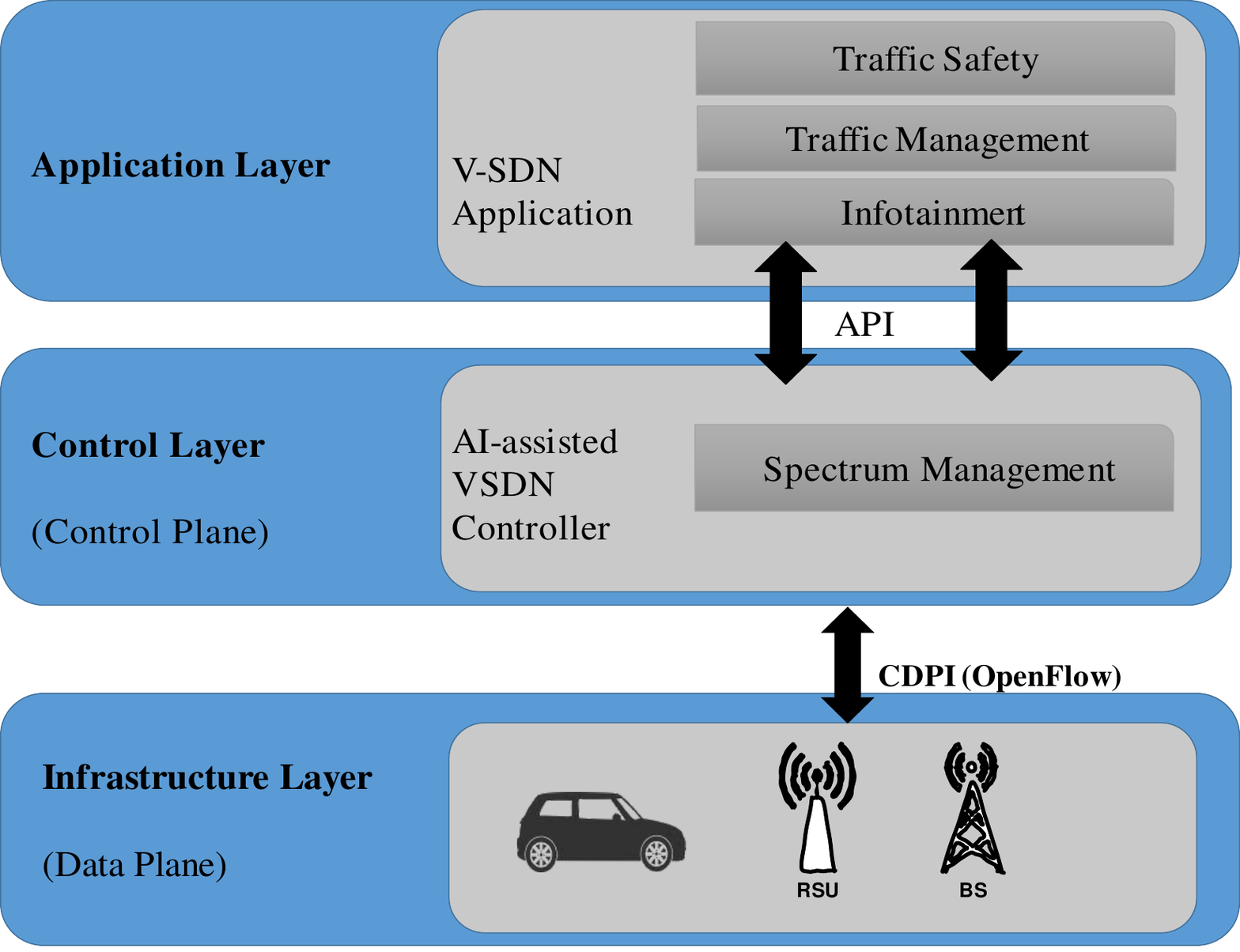}
    \caption{AI-assisted VSDN architecture.} 
    \label{fig:SpecVSDN}
\end{figure}

The potential for communication network congestion and the associated higher latency (due to queuing) are significant issues for latency-sensitive vehicular applications. 

Possible methods to help with these issues include multicarrier network access and software-defined heterogeneous vehicular networking architecture (e.g., SDN-based HetVNet~\cite{mahmood2019towards}). Multicarrier network access is a network selection technology that senses and selects the best of several available networks (e.g., DSRC or 5G) given these networks' properties and current congestion levels. A study on mobile latency on multiple operators in two distinct cities~\cite{multicarrierstudy} shows that such a carrier selection algorithm drops latencies 10 to 20\% compared to single carrier operations in real-time interactive cloud-based mobile applications such as augmented reality and cloud gaming. The technology also provides the potential for aggregation of several networks to increased bandwidth.

However, such solutions might not solve the congestion issue at the next hop (edge devices) or in the core network and might even exacerbate it. An edge-assisted Vehicular Software Defined Networking (VSDN) architecture could help solve this issue by installing VSDN controllers on edge devices at the network edge. The VSDN controllers can use continuously trained ML models to predict or detect periods of congestion and links with high utilization. They request more bandwidth upon detection of network bottlenecks by controlling the multi-carrier algorithm which either deactivates the current carrier and activates another one or activates another one and aggregates multi-carriers. The VSDN controllers also control the selection and aggregation onboard vehicles, and also find the best route or establish multiple connections to the destination node (e.g., cloud server) in the backhaul network to satisfy the bandwidth demand. Figure~\ref{fig:SpecVSDN} illustrates such an AI-assisted VSDN architecture. Vehicles and other road users can embed various connectivity modules (DSRC, LTE-A, Wi-Fi, mmWave) and select the most suitable RAT or combine two or more to ensure seamless and ubiquitous connectivity. In others words, the VSDN controller selects the communication channel with sufficient predicted capacity. 

\subsection{Privacy and Security challenges}
Remote vehicle diagnostics and maintenance, anomalous driving identification, and many other ITS applications are often based on learning from long-term data. They often require the transmission of potentially sensitive and private data (e.g., user identification, computational capacity, and insurance number ) to edge or cloud servers for data aggregation and computation offloading. To deal with this privacy issue, the automotive ecosystem should adopt federated learning (FL) for use cases where privacy is a serious concern and bandwidth is limited. Using an FL approach, users (e.g., vehicles, RSUs) share only locally trained model parameters which (compared to the raw data) are more difficult to extract insights from and are often smaller. Although FL copes with privacy and bandwidth, FL raises several new challenges such as model poisoning. Specifically, an attacker may poison the model by sending parameters of an anomalous model. 

Apart from FL, some vehicles (attackers) might initiate man-in-the-middle attacks to capture vehicle data (to attempt to learn about the drivers) or gain illegitimate remote control of another vehicle. It is thus necessary to define a set of policies and install tools to provide, confidentiality, reliability, integrity, and other security services. Vertical integration of security for data traffic from nearby vehicles and devices is critical at the node, domain, end-to-end, and service/use case levels. Besides, the integrity of transmitted and stored data is a crucial component in security provisioning. Meanwhile, a verifiable computing scheme (e.g., data attestation~\cite{chen2019deepattest}) for vehicular users is needed to check the correctness of any obtained computation results from the edge servers. Further research is needed to define the authorized users, vulnerabilities, and potential threats, and to create a trusted remote computing environment.

\subsection{Model Learning and Adaptation Over Time}
Driving patterns (along with other vehicular phenomena) are naturally dynamic and change over different timescales according to the road type, weather conditions, vehicle conditions, and driving style~\cite{martinez2017driving}. Thus, NN models, for example, for the prediction of driving patterns require training on significant heterogeneous historical data to account for these dynamics. However, given the possibility of rare novel events and non-stationarity, such models can benefit from continuous learning techniques (aka incremental learning). Specifically, these techniques allow learning from an online stream of incoming data (without full offline retraining) while avoiding the serious problem of forgetting previously learned data (known as catastrophic forgetting) by, for example, constraining how the network parameters can be updated during learning \cite{mai2022online}. Continuous learning approaches have not yet reached the performance levels of offline retraining and remain a major future research area.

Beyond continual learning, specific models are also designed for dynamic phenomena, for example, Gaussian-based dynamic probabilistic clustering (GDPC)~\cite{diaz2018clustering}. GDPC is a Gaussian mixture model-based unsupervised learning algorithm processing large amounts of data and coping with underlying dynamic phenomena (e.g., degradation). GDPC integrates three well-known algorithms: the expectation-maximization algorithm to estimate the model parameters, and the Page-Hinkley test and Chernoff bound~\cite{gama2014survey}. In turn, they use multiple (heterogeneous) data sources, fuse them, and based on which they train unsupervised or reinforcement learning models. If employed, these algorithms provide the model with the capability to detect the drift in the driving patterns.

\subsection{Augmented Reality}
Augmented reality (AR) windshields is a novel research area in both academia and industry that aims to improve driving safety and experience by augmenting environmental objects (e.g., roads, vehicles, obstacles, pedestrians) by overlaying helpful information. For example, a pedestrian image could be overlaid on the windshield at the location of an out-of-view pedestrian moving quickly towards the road, thus allowing the driver to be aware of the danger. However, AR windshields present significant challenges. Specifically, high motion-to-photon latency can cause misalignment between virtual objects and the physical world~\cite{liu2019align}, thus distracting the driver. Therefore methods to reduce networking and processing latency~\cite{7980118} including future communication technologies and computing paradigms as well as HCI methods to compensate for some degree of misalignment are important future research topics. Additionally, issues with the alignment of the driver's head with the windshield can cause other HCI issues. Finally, problems such as selecting which and how many objects to show or emphasize to the driver remain open.

\subsection{6G vision}
As mentioned, the widespread deployment of 5G vehicular communications will deliver, for example, higher data rates, lower latency, and more reliability to help support a variety of intelligent transportation systems. Research into 5G and 5G-advanced enabled ITS will continue to be a major research area with actual wide-scale deployment of such systems still years away. However, some future vehicular applications such as tactile internet use cases (where, for example, a remote operator would take control of an autonomous vehicle in an emergency) will require features 5G is lacking. Specifically, tactile internet requires sub 1 ms end-to-end latency and a combination of the several different 5G modes including ultra-reliable low latency and enhanced mobile broadband\cite{jiang2021road}. 6G vehicular communications look to support such applications that require multiple modes and overall aim to improve on most 5G KPIs by a factor of ten. Though, 6G also presents major research challenges such as maintaining reliability even with the use of high attenuation (yet large bandwidth) THz or optical wireless communication technologies that are potential major 6G components. Thus 6G vehicular communication will emerge as a significant future research topic in the coming years.

\section*{Acknowledgment}
\noindent This research has been supported in part by project 16214817 from the Research Grants Council of Hong Kong, and the 5GEAR and FIT projects from Academy of Finland.

\bibliographystyle{IEEEtran}
\bibliography{main}

\begin{IEEEbiography}[{\includegraphics[width=1in,height=1.25in,clip,keepaspectratio]{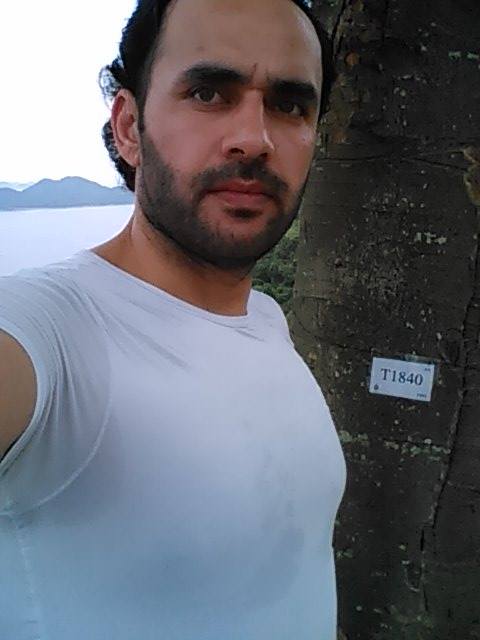}}]%
{Ahmad Yousef Alhilal}
 is currently pursuing Ph.D. degree at HKUST-DT Systems and Media Lab (SyMLab) in Computer Science Department, Hong Kong University of Science and Technology, Hong Kong. He received the bachelor's and M.S. degree from Damascus University, Syria. His research interests include vehicular communication and networking, edge computing, mobile cloud gaming, space communication and networking.
\end{IEEEbiography}
\begin{IEEEbiography}[{\includegraphics[width=1in,height=1.25in,clip,keepaspectratio]{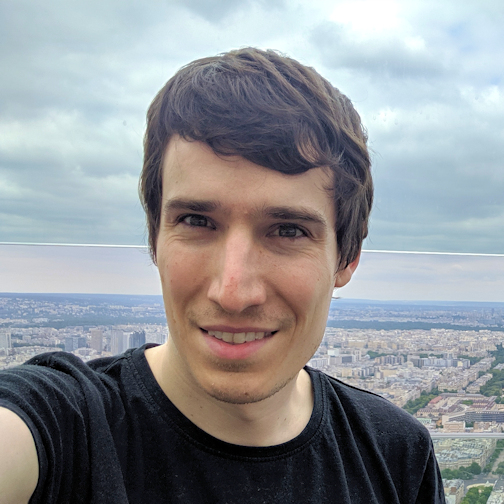}}]%
{Benjamin Finley} received the B.S. degree in software engineering from the Milwaukee School of Engineering, Milwaukee, WI, USA, and the M.S. and D.S. degrees in telecommunication engineering from Aalto University, Helsinki, Finland. He is currently a postdoctoral researcher with the Department of Computer Science, University of Helsinki. His current research interests include big telecom data analysis and user quality of experience.
\end{IEEEbiography}
\begin{IEEEbiography}[{\includegraphics[width=1in,height=1.25in,clip,keepaspectratio]{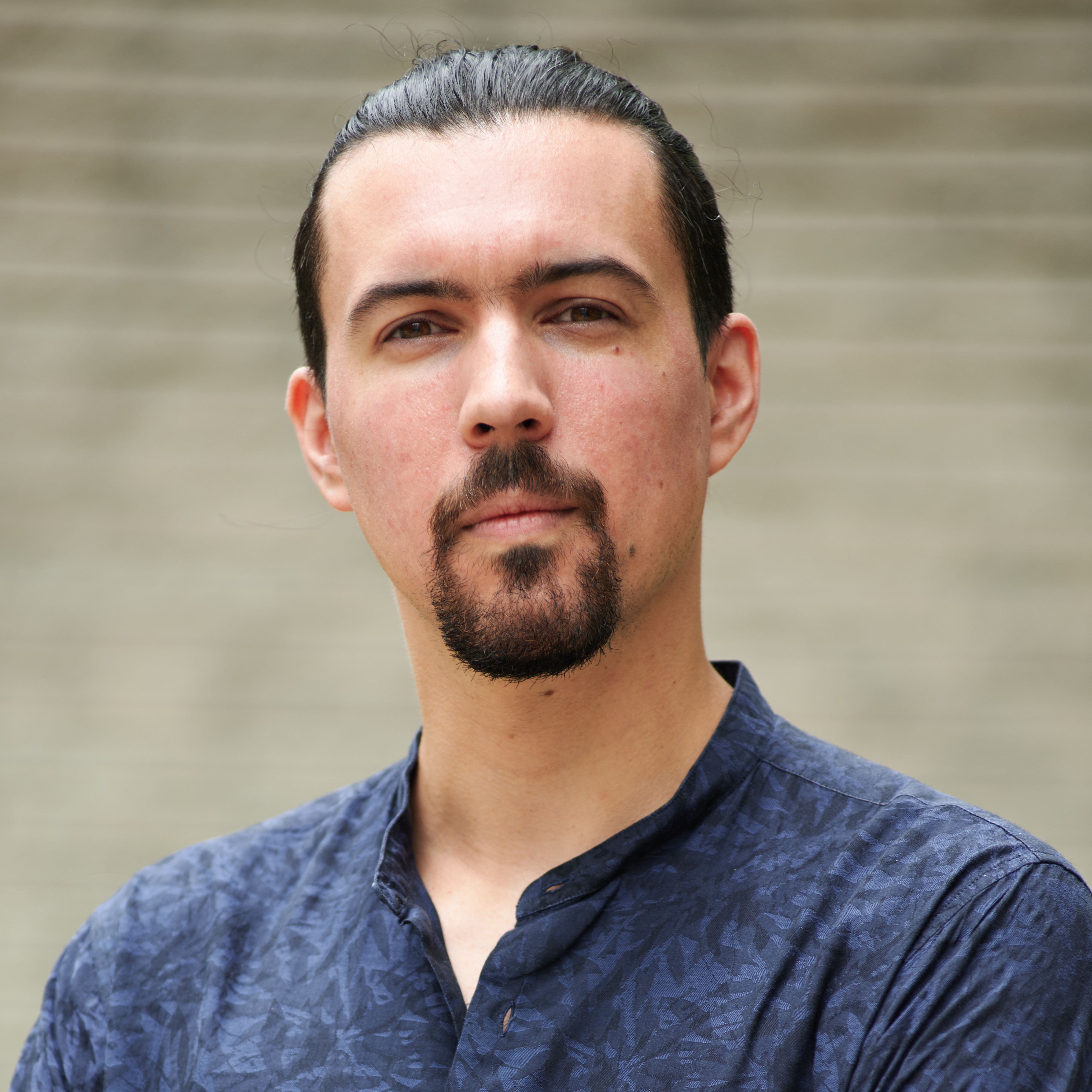}}]%
{Tristan Braud}
is an assistant professor at Division of Integrative Systems and Design, Hong Kong University of Science and Technology. He was a postdoctoral fellow at HKUST-DT Systems and Media Lab (SyMLab) in Computer Science Department, Hong Kong University of Science and Technology, Hong Kong. He got his Ph.D. degree from Université Grenoble Alpes, France in 2016. Before that, he was an engineering student at Grenoble INP Phelma, France, and received both a MSC from the Politecnico di Torino, Italy, and Grenoble INP, France. His major research interests include Augmented and Virtual Reality, with interests in pervasive and cloud computing as well as human centered system designs.
\end{IEEEbiography}
\begin{IEEEbiography}[{\includegraphics[width=1in,height=1.25in,clip,keepaspectratio]{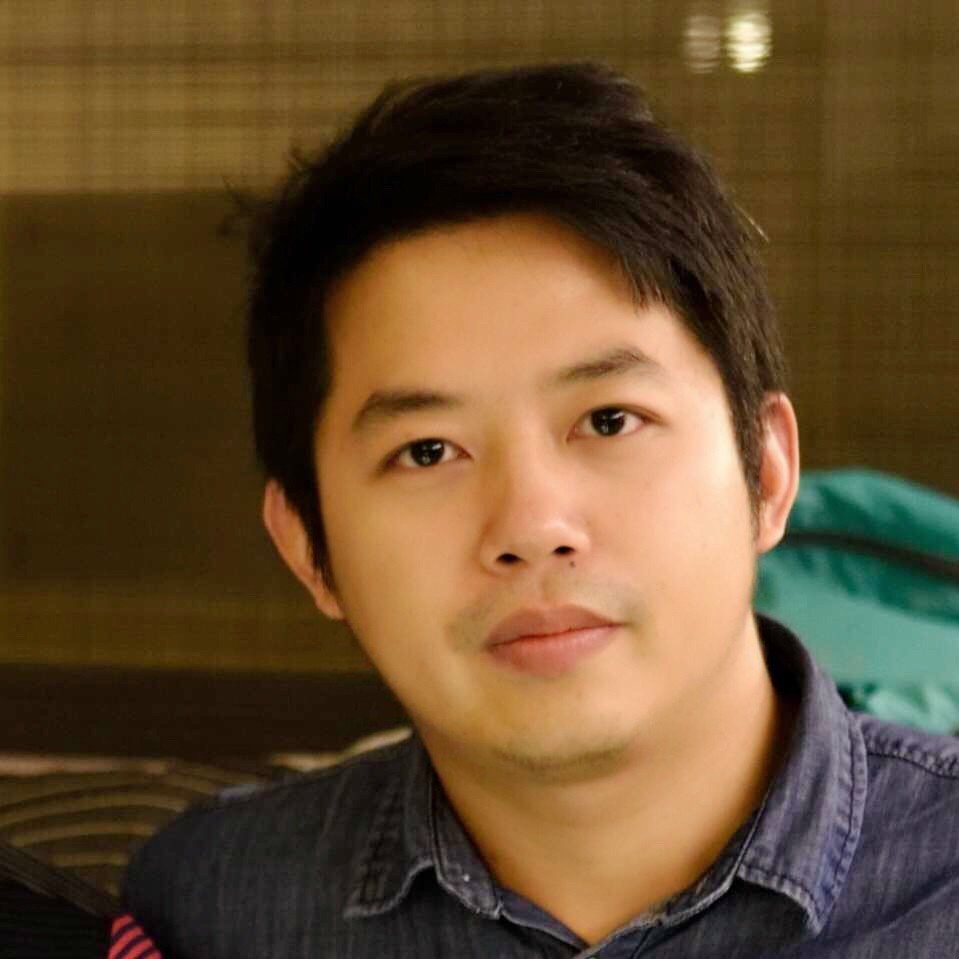}}]
{Dongzhe Su} 
is a Chief engineer in Communication Technologies Group of The Hong Kong Applied Science and Technology Research Institute. He has been leading the system architecture in R\&D of Vehicle-to-Everything (V2X) communication and application system, Connected Autonomous Vehicles (CAV) System, and Internet-of-Things (IoT). His role has been to define the technical scope and overall system design for ASTRI's V2X Networking system. He received MPhil degree in Computer Science from the Hong Kong University of Science and Technology. He received bachelor’s degree from Huazhong University of Science and Technology. In 2021, ASTRI launched one of the world's largest C-V2X public road tests in Hong Kong, covering a 14 km route with various road environment of Hong Kong. 
\end{IEEEbiography}
\begin{IEEEbiography}[{\includegraphics[width=1in,height=1.25in,clip,keepaspectratio]{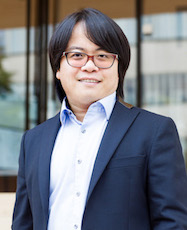}}]%
{Pan Hui}
(F’18) is a Professor of Computational Media and Arts and Director of the Center for Metaverse and Computational Creativity at the Hong Kong University of Science and Technology. He is also the Nokia Chair in Data Science at the University of Helsinki. He received his PhD from the Computer Laboratory at University of Cambridge, and both his Bachelor and MPhil degrees from the University of Hong Kong. He was an adjunct Professor of social computing and networking at Aalto University, Finland, and a Distinguished Scientist at the Deutsche Telekom Laboratories (T-labs), Germany. His industrial profile also includes his research at Intel Research Cambridge, UK and Thomson Research Paris, France. Pan Hui is a world-leading expert in Augmented Reality and Mobile Computing, with more than 400 research papers, 30 patents, and over 22,000 citations. He is an International Fellow of the Royal Academy of Engineering, a member of Academia Europaea, an IEEE Fellow, and an ACM Distinguished Scientist. He has founded and chaired many IEEE/ACM conferences/workshops, and has served as track chair, senior program committee member, organising committee member, and program committee member of numerous top conferences including ACM WWW, ACM SIGCOMM, ACM Mobisys, ACM MobiCom, ACM CoNext, IEEE Infocom, IEEE PerCom, IEEE ICNP, IEEE ICDCS, IJCAI, AAAI, UAI, and ICWSM. He served as an Associate Editor for IEEE Transactions on Mobile Computing and IEEE Transactions on Cloud Computing, and as guest editor for various journals including IEEE Journal on Selected Areas in Communications (JSAC) and IEEE Transactions on Secure and Dependable Computing. He also served on the IEEE Computer Society Fellow Evaluation Committee.
\end{IEEEbiography}
\EOD
\end{document}